\documentclass[aps,prl,twocolumn,amsmath,amssymb,superscriptaddress,showpacs]{revtex4-1}
\usepackage{graphicx}
\usepackage{dcolumn}
\usepackage{bm}
\usepackage{color} 
\definecolor{darkblue}{rgb}{0,0,0.6}
\definecolor{darkred}{rgb}{0.6,0,0}
\definecolor{darkgreen}{rgb}{0,0.6,0}
\usepackage[colorlinks=true,urlcolor=darkblue,citecolor=darkblue,linkcolor=darkred,hyperfootnotes=false]{hyperref}
\makeatother

\begin{document}

\title{Spontaneous Bending of Hydra Tissue Fragments Driven by Supracellular Actomyosin Bundles}

\author{Jian Su} 
\affiliation{Guangdong Technion - Israel Institute of Technology, 241 Daxue Road, Shantou, Guangdong, China, 515063}
\affiliation{Technion -- Israel Institute of Technology, Haifa, Israel, 3200003}

\author{Haiqin Wang} 
\affiliation{Guangdong Technion - Israel Institute of Technology, 241 Daxue Road, Shantou, Guangdong, China, 515063}
\affiliation{Technion -- Israel Institute of Technology, Haifa, Israel, 3200003}

\author{Zhongyu Yan} 
\affiliation{Guangdong Technion - Israel Institute of Technology, 241 Daxue Road, Shantou, Guangdong, China, 515063} 


\author{Xinpeng Xu}
\email{E-mail address: xu.xinpeng@gtiit.edu.cn}
\affiliation{Guangdong Technion - Israel Institute of Technology, 241 Daxue Road, Shantou, Guangdong, China, 515063}
\affiliation{Technion -- Israel Institute of Technology, Haifa, Israel, 3200003}

\date{\today}

\begin{abstract}

  Hydra tissue fragments excised freshly from Hydra body bend spontaneously to some quasi-stable shape in several minutes. We propose that the spontaneous bending is driven mechanically by supracellular actomyosin bundles inherited from parent Hydra. An active-laminated-plate model is constructed, from which we predict that the fragment shape characterized by spontaneous curvature is determined by its anisotropy in contractility and elasticity. The inward bending to endoderm side is ensured by the presence of a soft intermediate matrix (mesoglea) layer. The bending process starts diffusively from the edges and relaxes exponentially to the final quasi-stable shape. Two characteristic time scales are identified from the dissipation due to viscous drag and interlayer frictional sliding, respectively. The former is about 0.01 seconds, but the latter is much larger, about several minutes, consistent with experiments. 
  
\end{abstract}


\pacs{}

\maketitle

\begin{figure}[htbp]
  \centering
  \includegraphics[width=0.95\columnwidth]{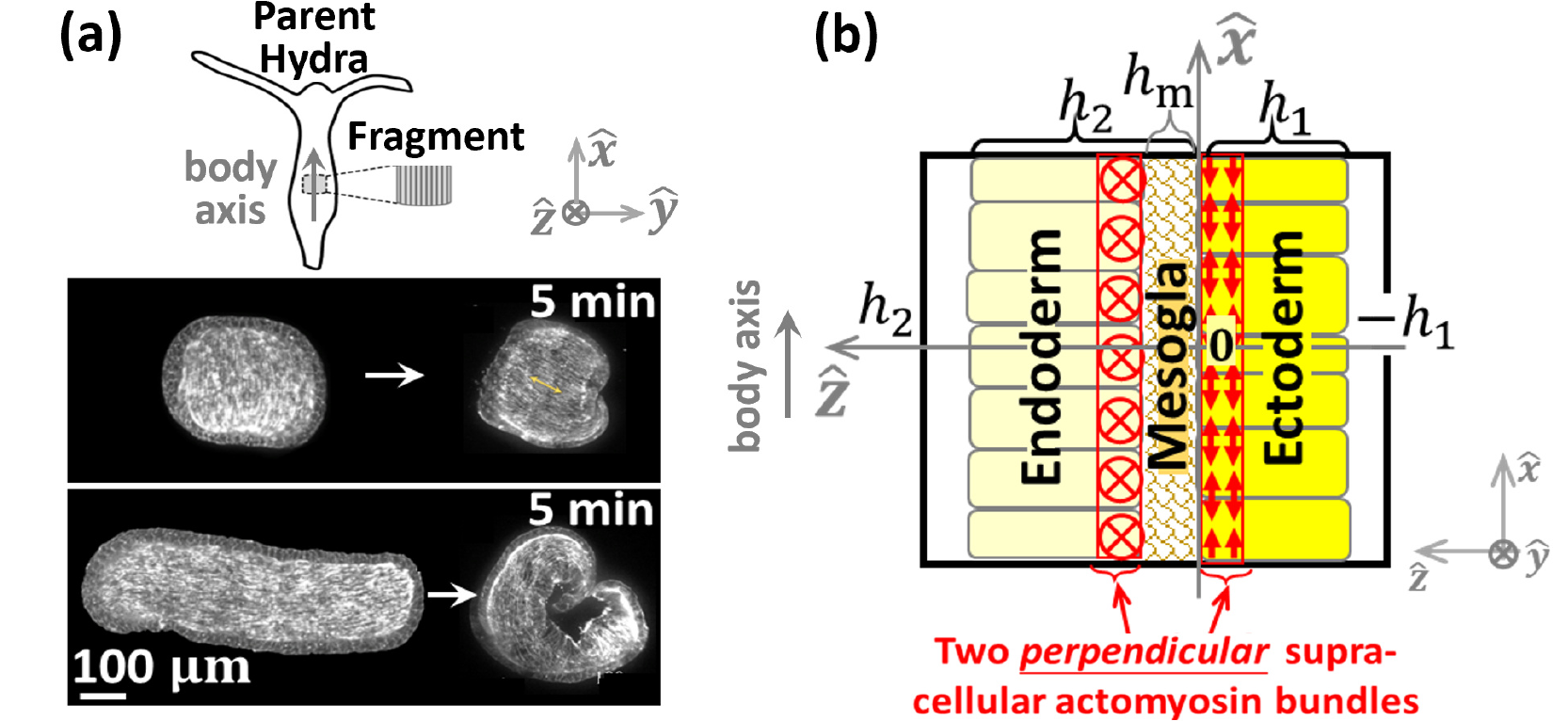}
  \caption {(color online). (a) Spontaneous bending of freshly excised Hydra (plate- and rod-like) tissue fragments to quasi-stable bent shape in several minutes.  
  (b) Triple-layer structure of Hydra fragments: ectoderm cell layer (yellow), endoderm cell layer (light yellow), and intermediate soft matrix (mesoglea). Two perpendicular supracellular actomyosin bundles (red arrows) are formed on the basal sides of each epithelial layer.
Reproduced from Livshits \emph{et al}.~\cite{Livshits2017} with permission from Elsevier.
  } \label{Fig:Schematic}
\end{figure}

Hydra is a multicellular fresh-water polyp, which exhibits remarkable regeneration capabilities, making it an excellent model system for studying tissue morphogenesis~\cite{Fujisawa2003,Takaku2005,Ott2008}. The Hydra body consists of a single-axis hollow-cylindrical tube (about $5-10 \, \rm{mm}$ long)~\cite{Fujisawa2003}, which has a triple-layer structure composed of two epithelial cell layers (endoderm and ectoderm) that are adhered together by an intermediate soft layer of extracellular matrix (called mesoglea) as shown in Fig.~\ref{Fig:Schematic}(b). Hydra body shape is maintained by a contractile actomyosin cytoskeleton~\cite{Livshits2017}, which shows a highly-aligned orientation over supracellular scales (see Fig.~\ref{Fig:Schematic}(b)): the actomyosin bundles are longitudinally oriented (along the tube axis) in the ectoderm and circularly oriented in the endoderm. Regenerating Hydras use their cytoskeleton to regulate their cells and to guide the regeneration process~\cite{Livshits2017}. When fragments are excised from the adult Hydra body, the supracellular pattern of cytoskeletal bundles are inherited, survive, and finally become a part of the new daughter Hydra. These supracellular cytoskeletal bundles provide structural ``memory" of the alignment along the body axis~\cite{Livshits2017}. They generate mechanical forces and direct the alignment of cytoskeleton in regions where the supracellular order is lacked in the early stage of regeneration. 


At the initial stage of the regeneration of Hydra tissue fragments that are excised freshly from adult Hydra body~\cite{Livshits2017,Krahe2013}, the fragments bend spontaneously to some quasi-stable (or mechanical-equilibrium) shape in several minutes (see Fig.~\ref{Fig:Schematic}(a)). Subsequently, the cytoskeleton of the fragment remodels to find a balance between maintaining its pre-existing organization and adapting to the new curved conditions. After about one hour of the bending-remodeling loop, the excised fragments fold into small hollow spheroids~\cite{Livshits2017,Krahe2013}.

\begin{figure}[htbp]
  \centering
  \includegraphics[width=0.95\columnwidth]{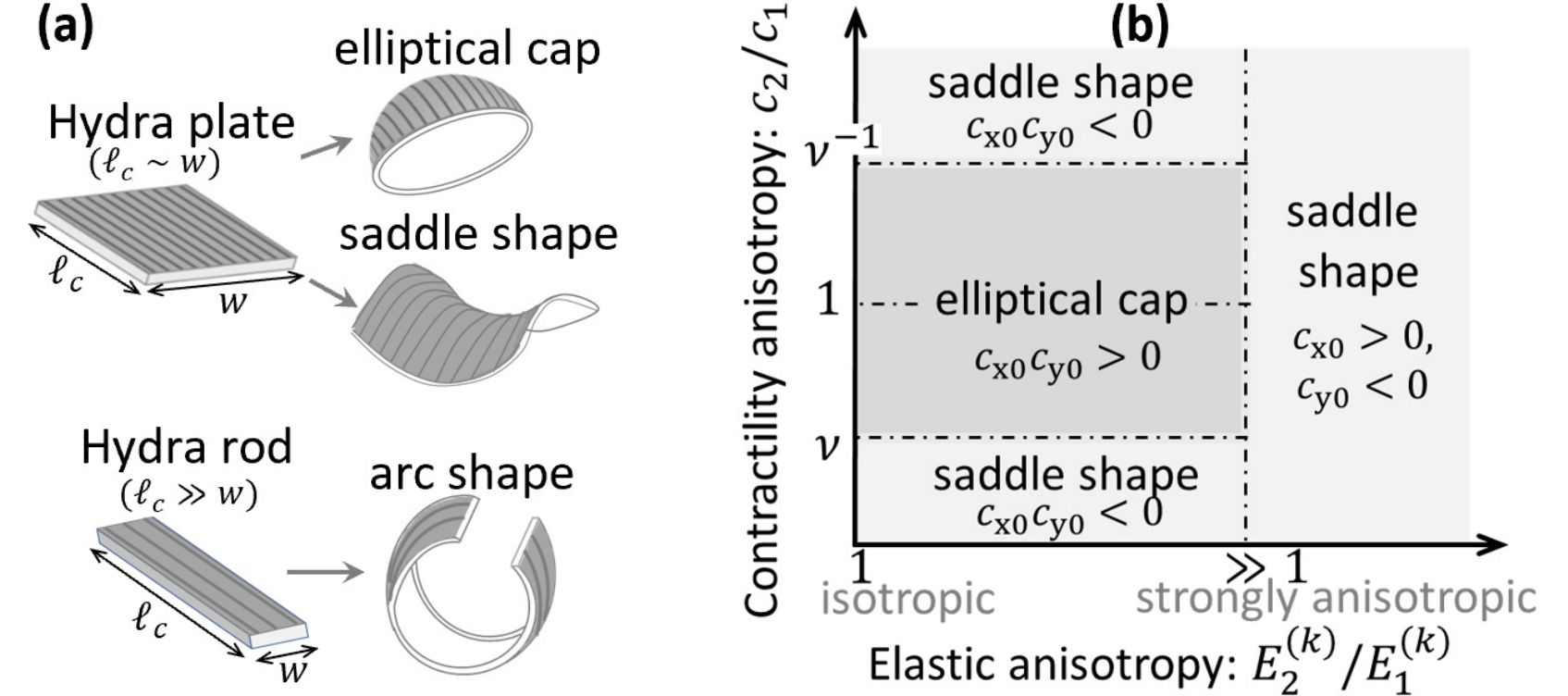}
  \caption {(a) Schematic illustration for the spontaneous bent shape of Hydra fragments. The contour length, width, and total thickness of the fragments are denoted by $\ell_{\rm c}$, $w$, and $h$, respectively. (b) A schematic bent-shape diagram of Hydra plates with different anisotropy in actomyosin contractility (measured by $c_2/c_1$) and in elasticity (measured by $E_2^{(k)}/E_1^{(k)}$). Here $c_k$ characterize the contractility of supracellular bundles; $E_1^{(k)}$ and $E_2^{(k)}$ are the two principal Young's moduli of each ($k=1,2$) orthotropic lamina layer.
  } \label{Fig:BentShape}
\end{figure}

In this Letter we focus on the initial spontaneous bending of Hydra fragments. Since the spontaneous bending happens very fast only in several minutes, it is reasonable to assume that the bending is accompanied by changes only in cell shape but not cellular rearrangements or cell division and apoptosis~\cite{Collins2016}. Furthermore, we propose that the spontaneous tissue bending is an active process driven mechanically by the contractility of the aligned, supracellular actomyosin bundles. An active-laminated-plate model is constructed that correlates the triple-layer structure and the supracellular contractile bundles with the spontaneous bending of Hydra fragments (see Fig.~\ref{Fig:Schematic}(b)). 
We predict that the bent shape of plate-like Hydra fragments characterized by the \emph{spontaneous curvature tensor} are mostly determined by their anisotropy in both the supracellular actomyosin contractility and the elastic properties of each lamina layer, as summarized in Fig.~\ref{Fig:BentShape}. 
The bending dynamics of rod-like Hydra fragments propagates \emph{diffusively} from the edges into the center (with the bent length being proportional to the square root of time). Whereas, when the bending is close to its final equilibrium (quasi-stable) shape, the end-to-end distance decays exponentially with time towards its equilibrium value. 
Moreover, we identify that the interlayer frictional sliding is the dominant dissipation mechanism governing the spontaneous bending of Hydra fragments in the first several minutes observed in experiments~\cite{Livshits2017}.

{\em Spontaneous curvature of Hydra tissue fragments. ---} To construct a continuum model for Hydra tissue fragments, we first note a structural analogy between Hydra fragments (with triple-layer structure) and composite laminated plates in material science~\cite{Landau1986,Reddy2003}. 
Moreover, we find that in typical regeneration experiments~\cite{Livshits2017}, the lateral lengths of Hydra fragments are usually $\sim 100 \, {\rm \mu m}$ much larger than their thickness $h \sim 10 \, {\rm \mu m}$. We then formulate a thin active laminated-plate (ALP) model~\cite{Xu2022,SM} for Hydra fragments by following the classical thin laminated plate theory~\cite{Reddy2003} that extends the classical thin plate theory~\cite{Landau1986} for mono-layer isotropic materials to multi-layer orthotropic materials. In our ALP model, the bent shape of a Hydra fragment is characterized by spontaneous curvature tensor and determined by minimizing the deformation energy of the laminate fragment upon internal active contraction applied by the two perpendicular supracellular actomyosin bundles. 

Technically, we treat the Hydra fragment as a laminated plate consisting of \emph{two} lamina layers (see Fig.~\ref{Fig:Schematic}(b)): the ectoderm layer-1, and the composite endoderm-mesoglea layer-2. In this case, we first consider Hydra fragments with isotropic lamina, where the total bending energy density (per area) is found to be~\cite{Xu2022,SM} 
\begin{align} \label{eq:Plates-Fb} 
F_{\rm b} = \frac{1}{2} D_{\rm p}(c_{\rm x}^2+2\nu c_{\rm x}c_{\rm y}+ c_{\rm y}^2) -(M_{\rm p}^{(1)} c_{\rm x}+M_{\rm p}^{(2)} c_{\rm y}), 
\end{align} 
with $c_{\rm x}$ and $c_{\rm y}$ being the two principal curvatures of the neutral surface at $z=z_0$. Here $\nu$ is the Poisson ratio and $D_{{\rm p}}$ is the effective flexural stiffness. $M_{\rm p}^{(k)}\approx \tau_{\rm p}^{(k)}z_0$ and $\tau_{\rm p}^{(k)}<0$ ($k=1,2$) are the active torques and contractile forces generated by supracellular actomyosin bundles in each of the two lamina layers, respectively. Note that contrasted with the bending energy in a quadratic form of principal curvatures in classical thin plate theory~\cite{Landau1986}, terms coupling active actomyosin torques linearly with curvatures~\cite{Sam2009,SamFred2012} also appear in Eq.~(\ref{eq:Plates-Fb}), which result in non-zero spontaneous curvatures. 

Minimizing $\mathcal{F}_{\rm b}=\int dxdy F_{\rm b} $ gives the principal spontaneous curvatures~\cite{SM}:
\begin{subequations}\label{eq:Plates-c0} 
\begin{equation}\label{eq:Plates-cx0cy0} 
c_{\rm x 0}=\frac{c_{1}-\nu c_{2}}{1-\nu^{2}}, \quad
c_{\rm y 0}=\frac{c_{2}-\nu c_{1}}{1-\nu^{2}},
\end{equation}
in which $c_k=M_{\rm p}^{(k)}/{D_{\rm{p}}}\approx{\tau_{\rm p}^{(k)}}z_0/{D_{\rm{p}}}$ ($k=1,2$) characterizes the contractility of the supracellular bundles in each lamina layer with the neutral surface position given by
\begin{align}\label{eq:Plates-z0} 
z_0=\frac{E_{\rm{p,eff}}^{(2)} h_{2}^{2}-E_*^{(1)} h_{1}^{2}}{2\left[E_*^{(1)}h_1+E_*^{(\rm m)}h_{\rm m}+E_*^{(2)} (h_2-h_{\rm m})\right]}. 
\end{align} 
\end{subequations}
Here $E_*^{(k)}\equiv E^{(k)}/(1-\nu^2)$; $E^{(k)}$ ($k=1,{\rm m},2$) are Young's modulus of the ectoderm, mesoglea, and endoderm, respectively. $h_{1}$ and $h_2$ are the thicknesses of the two lamina layers, respectively, with $h_{\rm m}$ being the thickness of mesoglea. The Poisson ratio $\nu$ is assumed to be constant ($0<\nu<1/2$) through the fragment thickness. 
$E_{\rm{p, eff}}^{(2)}=E_*^{(2)}-(E_*^{(2)}-E_*^{({\rm m})})h_{\rm m}^2/h_2^2$, is the thickness-averaged effective modulus of the endoderm-mesoglea layer-2. 
Note that the concept of spontaneous curvature has been used long time ago to represent the tendency of lipids to curve in lipid membranes~\cite{Sam2018} and to describe the spontaneous bending of uniformly-heated bimetal plates~\cite{Timoshenko1925}, in which the driving forces are chemical heterogeneity and asymmetric thermal expansion, respectively. In contrast, the spontaneous bending of tissue fragments proposed in this work are driven by completely different forces due to contractile supracellular actomyosin bundles.

From Eq.~(\ref{eq:Plates-c0}) we conclude that the spontaneous bending (with non-zero spontaneous curvatures) of Hydra fragments necessitates two basic conditions: 
the supracellular actomyosin contraction (represented by non-zero $\tau_{\rm p}^{(k)}$) and the internal asymmetry in the elastic properties and/or thicknesses of the two lamina layers (characterized by non-zero $z_0$). Moreover, for elastically isotropic fragments, we find from Eq.~(\ref{eq:Plates-cx0cy0}) that the principal curvatures and the equilibrium bent shape are mainly determined by the contractility anisotropy measured by $c_2/c_1$. If $\nu<c_2/c_1<\nu^{-1}$, the Hydra fragments will bend spontaneously to elliptical cap shape; otherwise, they will bend to saddle shape as shown in Fig.~\ref{Fig:BentShape}. 

Next, for tissue fragments consisting of strongly anisotropic lamina layers (for example, due to the presence of aligned thick actomyosin bundles in the two epithelial layers) we find that the spontaneous bending in the two principal directions are almost independent~\cite{SM}, and the neutral surfaces of the spontaneous bending lie in different layers for bending along different directions: in ectoderm and in endoderm for bending along the $\hat{\bf x}$-direction and $\hat{\bf y}$-direction, respectively. In this case, the two principal spontaneous curvatures are given by $c_{\rm x0}\approx c_1>0$, $c_{\rm y0}\approx c_2<0$, that is, the fragments bend spontaneously to be saddle shape (see Fig.~\ref{Fig:BentShape}(b)). 

Recent experiments~\cite{Livshits2017} show that (see Fig.~\ref{Fig:Schematic}(a)) Hydra fragments bend spontaneously to a spherical cap shape toward the inner endoderm side, that is, $c_{\rm x0} \approx c_{\rm y0}>0$ and $c_1 \approx c_2>0$. From the bent-shape diagram in Fig.~\ref{Fig:BentShape}(b), we find such cap-like bent shape corresponds to the case of more-or-less isotropy in both elasticity and contractility.  Moreover, in this case, we find from Eq.~(\ref{eq:Plates-c0}) that $z_{0}<0$, 
which requires $E_{\rm{p,eff}}^{(2)}<E_*^{(1)}$ (for $h_1\approx h_2$). This is ensured by the presence of the soft mesoglea layer with $E_*^{(\rm m)} \ll E_*^{(2)}$, which can reduce the effective endoderm-mesoglea modulus $E_{\rm{p, eff}}^{(2)}$
to $E_*^{(2)}(1-h_{\rm m}^2/h_2^2)$ that can be smaller than both the endoderm modulus $E_*^{(2)}$ and the comparable ectoderm modulus $E_*^{(1)}$~\cite{Collins2016}. Note that such mesoglea-softening mechanism is similar to that of connecting a stiff spring to a soft spring in series. Proper and stable inward tissue bending to endoderm side is essential for the formation of hollow Hydra spheroid and the whole Hydra regeneration process. 

{\em Spontaneous bending of rod-like Hydra fragments.---} If, instead of a Hydra plate, one cuts a rod-like Hydra fragment, the total bending energy density (per length) in Eq.~(\ref{eq:Plates-Fb}) reduces to~\cite{SM}: 
$F_{\rm b} = \frac{1}{2} D_{\rm r}c^2 -M_{\rm r} c$ with $c$ being the curvature of the neutral line at $z=z_0$. Here $D_{{\rm r}}$ is the effective flexural stiffness; $\tau_{\rm r} <0$ and $M_{\rm r} \approx \tau_{\rm r} z_0$ are the active contractile force and active torque generated by supracellular actomyosin bundle in the ectoderm layer-1, respectively. 
Minimization of $\mathcal{F}_{\rm b}=\int dx F_{\rm b} $ gives the spontaneous curvature~\cite{SM}: $c_0=M_{\rm r}/D_{\rm r} \approx \tau_{\rm r}z_0/D_{\rm r}$, and the neutral line position $z_0$ is given by the same form as Eq.~(\ref{eq:Plates-c0}) if replacing $E_*^{(k)}$ by $E^{(k)}w$ with $w$ being the width of the Hydra rod. 

Using $D_{\rm r}\sim E h^3 w$, $z_0 \sim h$, and noting that in typical cell experiments~\cite{Collins2016,Prost2021}: $\tau_{\rm r}/h w\sim 0.1 \,\rm{kPa}$, $h \sim 10 \,\rm{\mu m}$, $E\sim 1 \,\rm {kPa}$, we estimate the spontaneous curvature $c_0 \sim 1/100 \, \rm{\mu m}^{-1}$. Interestingly, for Hydra regenerating from cell aggregates~\cite{Ott2008}, the curvature $c$ of the hollow Hydra spheroids formed from an aggregate of a minimal number of $N_{\rm c} \sim 1000$ cells~\cite{Takaku2005,Ott2008} can be estimated to be the same scale $\sim 1/100 \, \rm{\mu m}^{-1}$ by calculating the surface area of the spheroid, $4\pi c^{-2} \sim N_{\rm c} R_{\rm c}^2$, and using the cell size $R_{\rm c}\sim 10\, \rm{\mu m}$. Moreover, such cell-scale curvature is also known to appear very often in many natural \emph{in vivo} microenvironment~\cite{Sam2009}, \emph{e.g.}, cylindrical shaped glands and blood vessels~\cite{Callens2020}. 


Note that the supracellular contraction or the internal asymmetry in elastic and/or geometric properties are generally non-uniform and hence the spontaneous curvature can be different at different positions through the fragments. In this case, some non-trivial periodic bent shapes have been obtained as shown in Fig.~S2 of SM~\cite{SM}. 
Moreover, we would like to point out that the mechanistic perspective of the spontaneous bending of tissue fragments during morphogenesis conveyed in this work has been proposed before~\cite{Siber2017,Krahe2013,Joanny2014,Ackermann2022} and several classical theories have been developed. Among the earliest theories of this kind, Lewis proposed (1947)~\cite{Siber2017} a mechanical model of epithelial sheets, consisting of brass bars, tubes, and rubber bands. We discuss this model in SM~\cite{SM} and compare it with our ALP model for the spontaneous bending shape of Hydra tissue.

In addition, during the tissue bending, interlayer slide may occur between each cell layer and the intermediate mesoglea matrix~\cite{Prost2021}, and the fragment becomes incoherent~\cite{Sam1994}. In this case, we neglect the in-plane contraction and the total energy $\mathcal{F}_{\rm t}$ takes the following phenomenological form~\cite{Sam1994,SamFred2012} 
\begin{align}\label{eq:Rods-Ft0}
{\cal F}_{\rm t}= \int_{-\ell_{\rm c}/2}^{\ell_{\rm c}/2} ds \left[\frac{Y_{\rm s}}{2}\epsilon_{\rm s}^2 -\chi \epsilon_{\rm s} c+ \frac{{D}_{\rm r}}{2} \left(c-\tilde{c}_{0}\right)^{2}\right].
\end{align}
Here $\ell_{\rm c}$ is the rod contour length, $\epsilon_{\rm s}$ denotes the interlayer slide or the misfit strain discontinuity between the two lamina layers, and $Y_{\rm s}$ is the stiffness for the interlayer slide. $\tilde{c}_{0}$ being the spontaneous curvature of coherent fragments. $\chi$ is the linear coupling coefficient. The equilibrium bent state of the rod is then characterized by the spontaneous curvature $c_0\equiv \tilde{c}_{0}/{(1-\chi^2/D_{\rm r}Y_{\rm s})}$ and the equilibrium slide $\epsilon_{\rm s0}=\chi c_0/Y_{\rm s}$. Particularly, in the limit of $Y_{\rm s}\to \infty$, we recover the case of coherent fragments with $c_0= \tilde{c}_{0}$ and $\epsilon_{\rm s0}=0$ where the strain is continuous through the thickness.

{\em Bending dynamics of tissue rods: effects of viscous dissipation and interlayer frictional sliding. ---} We now consider the dynamic process of the spontaneous bending of Hydra tissue rods in surrounding viscous fluids starting from initial flattened state to the final equilibrium shape as shown in experiments (Fig.~\ref{Fig:Schematic}(a)). 
The bending dynamics of long soft rods with non-zero spontaneous curvature $c_0$ and contour length $\ell_{\rm c}\gg 2\pi R_0$ ($R_0=c_0^{-1}$) has been explored intensively in both theory and experiments~\cite{Brochard2012,Callan2012}. However, in typical Hydra regeneration experiments~\cite{Livshits2017}, Hydra rods are usually short with $\ell_{\rm c} \sim 2\pi R_0$. In this case, the total energy $\mathcal{F}_{\rm t}$ of the laminated Hydra rod is given by Eq.~(\ref{eq:Rods-Ft0}), and the bending dynamics is driven by the relaxation of stored elastic energy in the initial flattened state. Furthermore, the dissipation during the bending arises mainly from viscous drag in the surrounding fluids and the interlayer frictional sliding inside the fragments, as taken into account by the dissipation function
\begin{equation}\label{eq:BendDyn-Phi} 
\Phi=\int_{-\ell_{\rm c}/2}^{\ell_{\rm c}/2} ds \left(\frac{1}{2} \xi_{\rm v} \dot{\bm{r}}^{2}+\frac{1}{2} \xi_{\rm s} \dot{\epsilon}_{\rm s}^{2} \right).
\end{equation}
Here $\bm{r}(\rm s)$ is the position vector of points (parameterized by arc length $\rm s$) on the rod, and the local curvature $c$ is given by its second derivatives~\cite{Sam2018}. $\xi_{\rm v}$ and $\xi_{\rm s}$ are the viscous drag coefficient per rod length and friction coefficient for interlayer sliding, respectively. 
Interestingly, the interlayer frictional sliding has recently been found to be able to facilitate the long-range force propagation in tissues that usually have multi-layer structures~\cite{Prost2021}. 

We then use Onsager's variational principle~\cite{Doi2021,Xu2021} and minimize Rayleighian ${\cal R}[\dot{\bm{r}}, \dot{\epsilon}_{\rm s}] = \dot{\cal F}_{\rm t}+\Phi$ yielding a set of highly nonlinear partial differential equation (\emph{e.g.}, Kirchhoff equations)~\cite{Callan2012}. However, here we will not solve these nonlinear equations but carry out direct-variational analysis~\cite{Doi2021,Xu2021} for the very initial bending stage near the flattened state and the final bending stage close to the equilibrium bent shape.

\begin{figure}[htbp]
  \centering
  \includegraphics[width=0.95\columnwidth]{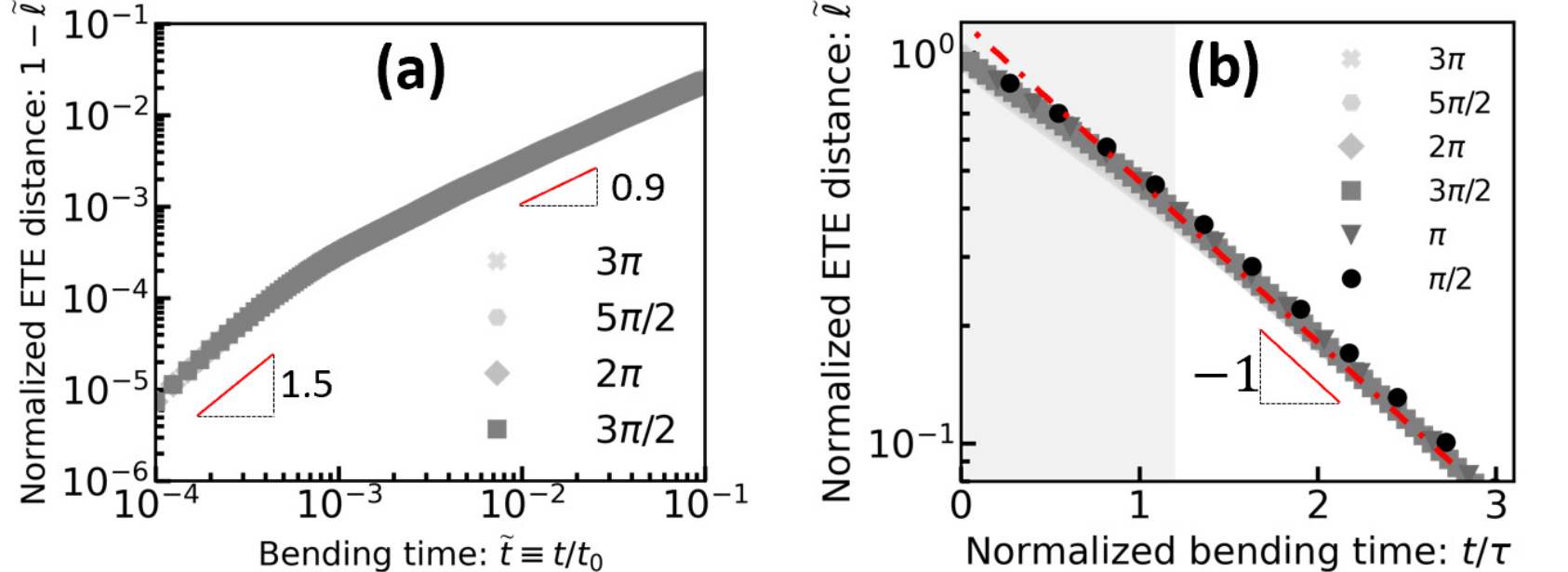}
  \caption {Temporal evolution of the end-to-end (ETE) distance $\tilde{\ell}$ for Hydra rods: (a) the initial bending starting diffusively from the edges, and (b) the final bending relaxing exponentially to the equilibrium shape. Universal scaling relations are found for rods of different contour lengths $\pi/2 \leq \tilde{\ell}_{\rm c} \leq 3\pi$. 
  } \label{Fig:BendDyn}
\end{figure}

Initially near the flattened state, we neglect the effects of interlayer slide and take a linear trial profile of local curvature (see Fig.~S4(a) in SM~\cite{SM}), where the dynamic bending process is characterized by one time-dependent parameter: the bent length $a(t)$ of the rod. 
Substituting the trial profile into Eqs.~(\ref{eq:Rods-Ft0}) and (\ref{eq:BendDyn-Phi}), we obtain the Rayleighian $\mathcal{R}\approx -\frac{1}{3}D_{\rm r}c_0^2 \dot{a}+\frac{1}{2}\xi_{\rm v} a \dot{a}^2$ and its minimization with respect to $\dot{a}$ gives
\begin{equation}\label{eq:at}
a \approx \left(\frac{2 D_{\rm r} c_{0}^{2}}{3\xi_{\rm v}}\right)^{1/2} t^{1/2},
\end{equation}
that is, the bending starts from the edges and propagates \emph{diffusively} into the center. Moreover, from Eq.~(\ref{eq:at}) we calculate the normalized end-to-end (ETE) distance $\tilde{\ell} \equiv ({\ell-\ell_{\rm eq}})/({\ell_{\rm c}-\ell_{\rm eq}})$ scaling as $1-\tilde{\ell} \sim a^3 \sim  t^{3/2}$, where $\ell$ is the ETE distance with $\ell_{\rm eq}$ being its equilibrium value. Interestingly, although the linear trial profile has some visible deviation from the simulated profile (as shown in Fig.~S4(a) in SM~\cite{SM}), the scaling law predicted above for ETE distance agrees well with simulation results in Fig.~\ref{Fig:BendDyn}(a) and is not sensitive to the form of trial curvature profile; the same scaling is obtained by using a simpler trial step-function profile of curvature. However, we note that the initial diffusive bending process happens only in a very short period as shown in Fig.~\ref{Fig:BendDyn}(a). The bending dynamics slows down quickly from the diffusive regime to a subdiffusive regime with $1-\tilde{\ell} \sim  t^{0.9}$. 

In the final stage of the bending dynamics close to the quasi-stable bent state, we take the same strategy as above and also choose a linear trial profile of the curvature (see Fig.~S4(c) in SM~\cite{SM}), where the dynamic bending process is described by the two time-dependent parameters: the normalized curvature $\tilde{c}_{\rm m}(t)=c_{\rm m}/c_0$ and interlayer slide $\tilde{\epsilon}_{\rm s,m}(t)=\epsilon_{\rm s, m}/\epsilon_{\rm s0}$ in the rod center at $s=0$.  
Substituting the trial profile into Eqs. (\ref{eq:Rods-Ft0}) and (\ref{eq:BendDyn-Phi}), we obtain the Rayleighian $\mathcal{R}$ and its minimization yields
\begin{align} \label{eq:BendDyn-regime1-DynEqn}
\tau \dot{\tilde{c}}_{\rm m}=1-(1+{\cal B}) \tilde{c}_{\rm m}+{\cal B} \tilde{\epsilon}_{\rm s,m}, \quad \tau_{\rm s} \dot{\tilde{\epsilon}}_{\rm s,m}=\tilde{c}_{\rm m} -\tilde{\epsilon}_{\rm s,m}, 
\end{align}
where ${\cal B}={Y_{\rm{s}} \epsilon_{\rm s0}^{2}}/{\tilde{D}_{\rm{r}} c_{0}^{2}}$ characterizes the relative stiffness of interlayer slide and out-plane bending with $\tilde{D}_{\rm{r}} \equiv {D}_{\rm r}-\chi^2/Y_{\rm{s}}$ being the normalized flexural stiffness, $\tau =6 {\xi_{\rm v}} \mathcal{L} /{\tilde{D}_{\rm{r}} c_0^4}$ and $\tau_{\rm s} = {\xi_{\rm s}}/{Y_{\rm{s}}}$ are the two characteristic time scales associated with viscous drag and interlayer frictional sliding, respectively. 
Note that $\mathcal{L}$ is a nonlinear function of rod contour length $\ell_{\rm c}$ (changing from $0$ to order one; see its expression in SM~\cite{SM}). 


We consider the following two limits of the dynamic equation~(\ref{eq:BendDyn-regime1-DynEqn}) that are particularly interesting and may be relevant to the bending dynamics during tissue morphogenesis and regeneration. 
Firstly, in the limits of either coherent Hydra rods or incoherent Hydra rods with small friction, we have $\tau_{\rm s}/\tau \ll 1$ and the viscous drag from the surrounding fluids is the dominant dissipation mechanism. In this limit, we find from Eq.~(\ref{eq:BendDyn-regime1-DynEqn}) the normalized ETE distance $\tilde{\ell} (t)$ follows $\tilde{\ell} \sim \exp\left(-t/\tau\right)$, 
which have been justified for rods with various contour lengths by numerical simulations using bead-spring model and shown in Fig.~\ref{Fig:BendDyn}(b). 
In addition, we have used bead-spring model to numerically examine the bending dynamics of Hydra rods under some different boundary conditions (see Figs.~S5-S7 in~\cite{SM}). We find that the bending dynamics predicted above is very robust: the same scaling laws (including the nonlinear contour-length dependence of relaxational time $\tau$ due to $\mathcal{L}$) are found for most different boundary conditions in both the initial diffusive bending and the final relaxational bending processes.



Secondly, in the limit of incoherent Hydra rods with large friction, we have $\tau_{\rm s}/\tau \gg 1$, we obtain from Eq.~(\ref{eq:BendDyn-regime1-DynEqn}) the ETE distance $\tilde{\ell} (t)$ following the same exponential form:
\begin{equation}\label{eq:BendDyn-regime2-ETEt}
\tilde{\ell}\sim \exp\left[-t/\tau_{\rm s}(1+{\cal B})\right],   
\end{equation}
in which  ${\cal B}\sim 1$ so that all energy terms in Eq.~(\ref{eq:Rods-Ft0}) are comparable.
That is, in either limits, the characteristic time for the spontaneous bending of tissue rod is controlled by the slower dissipative dynamics and the longer time scale. Particularly for Hydra tissue rods, we can estimate the magnitude of the two time scales as follows. 
Using $\tilde{D}_{\rm{r}}\sim E h^3 w$, $\xi_{\rm v} \sim \eta$, $w\sim h $, $Y_{\rm{s}} \sim Eh w$, and $\xi_{\rm s}=\xi h^2 w$ with $\xi$ being the interlayer friction coefficient~\cite{Prost2021}, we obtain 
\begin{equation} 
\tau \sim \frac{\xi_{\rm v}}{\tilde{D}_{\rm{r}} c_0^4} \sim \frac{\eta}{E h^4 c_0^4} \sim 0.01 \, {\rm s }, \quad
\tau_{\rm s} \sim \frac{\xi h}{E}  \sim 1 \, {\rm min},
\end{equation}   
where we have taken typical parameter values measured in cell experiments~\cite{Collins2016,Prost2021}: $h \sim 10\, {\rm{\mu m}}$, $c_0 \sim 1/100 \, {\mu m}^{-1}$, $\eta \sim 10^{-3} \,\rm{Pa\cdot s}$, $E \sim 1 \, \rm{ k Pa}$,  and $\xi \sim 10^{10} \, \rm{Pa\cdot s/m}$. Therefore, the spontaneous bending of Hydra rods lies in the limit of $\tau_{\rm s}/\tau \gg 1$ and is characterized by $\tau_{\rm s} \sim 1 \, {\rm min}$ in consistent with the time duration of initial spontaneous bending observed in Hydra regeneration experiments (see Fig.~\ref{Fig:Schematic}(a))~\cite{Livshits2017}. 

The proposal here on the spontaneous bending of Hydra tissue fragments driven by contractile supracellular actomyosin bundles should be generically present in a broad range of cell aggregates and tissue fragments during tissue regeneration or morphogenesis during embryo development. Our mechanical model of Hydra fragments can be easily extended to study the morphogenesis of other tissues and to include the couplings between the organization of cytoskeleton and the deformation/curvature of the tissue~\cite{Len2020}. The effects of the permeation of water or other small molecules on the dynamics of tissue bending can be explored similarly by modeling the tissue fragments as active-laminated gels~\cite{Julicher2011,Doi2022}. Additional aspects of the cell biology, such as cell division/apoptosis and cell morphological changes~\cite{Joanny2014,Ackermann2022}, or active behaviors such as migration and oscillations, could be incorporated as well~\cite{Siber2017,Julicher2017}. 

%
%

The authors thank Samuel Safran, Kinneret Keren, Zhihao Li, Yonit Maroudas-Sacks, Lital Shani-Zerbib, and Anton Livshits for fruitful discussions. X.X. is supported in part by National Natural Science Foundation of China (NSFC, No.~12004082), by Guangdong Province Universities and Colleges Pearl River Scholar Funded Scheme (2019), and by 2020 Li Ka Shing Foundation Cross-Disciplinary Research Grant (No.~2020LKSFG08A). J. Su and H. Wang contributed equally to this work.

\newpage

\newpage
\bibliography{HydraFolding}

\end{document}


\title{Supplementary Material for ``Spontaneous Bending of Hydra Tissue Fragments Driven by Supracellular Actomyosin Cables''}
\date{\today}

\pacs{}
\maketitle

\tableofcontents

~\\

In this supplementary material, we provide more details of our computations and discussions about the statics and dynamics of the spontaneous bending of Hydra tissue fragments. In Sec. I, we first calculate the spontaneous curvature of Hydra fragments by modeling the fragments as composite active laminated plates (ALP) with asymmetric internal (active cellular) contraction. We address the importance of anisotropy in cellular contractility and elastic properties in determining the bent shape of Hydra fragments. Particularly, for rod-like fragments we show that our composite ALP model of tissue fragments has a simple mapping to the classical brass-bar-and-rubber-band model for tissue morphogenesis that was proposed by W. H. Lewis in 1947. In addition, we have also discussed the effects of inter-layer sliding briefly. In Sec. II, we present our investigations on the bending dynamics of short Hydra rods with non-zero spontaneous curvature. Two dynamic regimes have been identified from both approximate analytical calculations based on Onsager's variational principle and numerical simulations using the bead-spring model of thin rods. Particularly, we predict that the bending of coherent rods upon viscous drag in surrounding fluids usually happens very fast in the time scale of $0.01$ seconds. However, the presence of inter-layer frictional sliding in incoherent Hydra rods can slow down the bending dynamics significantly and the characteristic bending time becomes the order of magnitude of several minutes.  

\section{Spontaneous curvature of Hydra tissue fragments \label{sec:c0}}

\subsection{Hydra tissue fragments modelled as composite active laminated plates (ALP) \label{sec:c0-Plates}}

\begin{figure}[htbp]
  \centering
  \includegraphics[width=0.8\columnwidth]{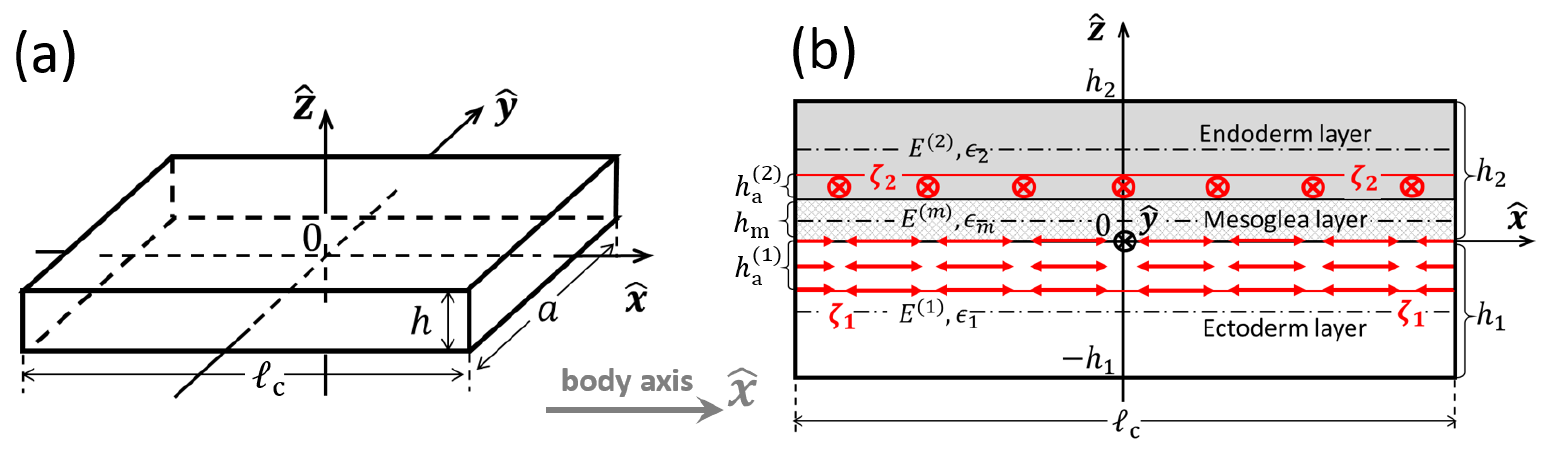}
  \caption {Schematic illustrations for the composite active-laminated-plate (ALP) model of Hydra tissue fragments upon internal cellular contraction. (a) Illustration of the 3D Cartesian coordinates of thin composite active laminated plates. Body axis of the Hydra is along the $+x$-direction. (b)  Illustration of the triple-layer structure of Hydra tissue fragment, consisting of the ectoderm, endoderm, and  mesoglea with thicknesses, $h_1$, $h_2$, and $h_{\rm m}=\alpha_{\rm m}h_2$, respectively. Here two perpendicular supracellular layers contracting uniaxially (along $x-$ and $y-$ directions respectively) appear in the two cell layers, respectively. The thickness and the active stress of the two contracting layers are $h_{\rm a}^{(1)}\equiv \alpha_1 h_1$, $h_{\rm a}^{(2)}\equiv \alpha_2 h_2$, and ${\bm \sigma}^{\rm a (1)} =\zeta_1 \hat{\bf x}\hat{\bf x}$, ${\bm \sigma}^{\rm a(2)} =\zeta_2 \hat{\bf y}\hat{\bf y}$, respectively. Note that $E^{(k)}$ denotes the elastic modulus of the $k-$th layer. 
  } \label{Fig:Schematic-Plates}
\end{figure}

Hydra tissue fragments cut from adult Hydra body have a triple-layer structure consisting of two epithelial cell layers (endoderm and ectoderm) that are separated by an intermediate layer of  extracellular matrix (gel-like substance, called mesoglea), as show in Fig. 1 of the main text. During the initial spontaneous bending of a Hydra tissue fragment, the Hydra tissue fragment can be modeled simply as an elastic composite (triple-layer) laminate plate and the spontaneous bending is driven by the two perpendicularly-oriented supracellular actomyosin cables in the basal sides of the two cell layers: the longitudinally oriented cables in the ectoderm and the circularly oriented cables in the endoderm. As schematically shown in Fig.~\ref{Fig:Schematic-Plates}, the Hydra-body axis is taken to be along the $x-$direction and the two perpendicular supracellular actomyosin cables in the ectoderm and endoderm are aligned along the $x-$ and $y-$directions, respectively, as shown in Fig.~\ref{Fig:Schematic-Plates}(b).  

Such trilayer structure of Hydra tissue fragments are similar to that of composite laminates with orthogonal fiber reinforcement in material science \cite{Lekhnitskii1981,Reddy2017}. Here in formulating the theory for the spontaneous bending of Hydra tissue fragments, we follow the classical laminated plate theory (CLPT) \cite{Reddy2003,Xu2022} and propose that the spontaneous curvature of a Hydra tissue fragment is determined by minimizing the deformation energy of the laminated-plate fragment upon internal active cellular contraction. In this case, we make the following assumptions or restrictions~\cite{Reddy2003,Xu2022}: 
\begin{itemize}
\item Each lamina layer is of uniform thickness. The thicknesses of the ectoderm, endoderm, and mesoglea are denoted by $h_1$, $h_2$, and $h_{\rm m}=\alpha_{\rm m}h_2$, respectively, with $\alpha_{\rm m}$ being the fraction of mesoglea layer in the composite (endoderm and mesoglea) lamina layer-2. 
\item Each lamina layer is a linear orthotropic elastic material (with \emph{nine} independent elastic constants) \cite{Landau1986,Reddy2017}. One symmetry ($xy$) plane is parallel to the cell layers, the other two symmetry planes ($xz$ and $yz$) are perpendicular to the two supracellular actomyosin cables, respectively. 
\item The lamina layers are perfectly bonded together (that is, coherent). Note that this restriction is resolved when we consider the effects of inter-layer sliding. 
\item The Hydra tissue fragments are thin since their thickness $h$ (in the $z$ direction with $h=h_1+h_2+h_{\rm m}$) are usually much smaller than their dimensions in the other two lateral ($x,\,y$) directions \cite{Landau1986,Reddy2017} (as shown in Fig.~\ref{Fig:Schematic-Plates}(a)).

\item The displacements and strains of the Hydra tissue fragments are small.  
\item The two supracellular actomyosin cables are mimicked by two perpendicular contracting layers of small thickness, $h_{\rm a}^{(1)}\equiv \alpha_1 h_1$ and $h_{\rm a}^{(2)}\equiv \alpha_2 h_2$, in the ectoderm (layer 1) and the endoderm (layer 2), respectively (see Fig.~\ref{Fig:Schematic-Plates}(b)). Then the active stress tensor that depends on $z$ can be written as 
\begin{subequations}\label{eq:Plates-hsigmaa}
\begin{equation}\label{eq:Plates-hsigmaa1}
{\bm \sigma}^{\rm a} = \zeta^{(1)} \hat{\bf x}\hat{\bf x}+\zeta^{(2)} \hat{\bf y}\hat{\bf y},
\end{equation} 
in which the components $\zeta^{(1)}$ and $\zeta^{(2)}$ follow simple step-wise forms as
\begin{align}\label{eq:Plates-zeta} 
\zeta^{(1)} = 
\begin{cases} 
\zeta_1,  &\mbox{if} \quad -h_{\rm a}^{(1)}<z<0, \\
0, & \mbox{if} \quad \rm{otherwise},
\end{cases}
\quad 
\zeta^{(2)} = 
\begin{cases} 
\zeta_2,  &\mbox{if} \quad h_{\rm m}<z<h_{\rm m}+h_{\rm a}^{(2)}, \\
0, & \mbox{if} \quad \rm{otherwise},
\end{cases}
\end{align}
\end{subequations}
with $\zeta_1,\, \zeta_2<0$ for contractile stresses. 
\end{itemize}  

\subsubsection{Total deformation energy of thin active laminated plates in the CLPT } \label{sec:c0-Plates-general}

In the classical laminated plate theory (CLPT) for thin laminates, the Kirchhoff hypothesis is assumed to hold, which amounts to neglecting both transverse shear and transverse normal effects, and the transverse stresses and transverse shear strains are identically zero: 
\begin{equation}
\sigma_{\rm zz}=\sigma_{\rm xz}=\sigma_{\rm yz}=0, \quad
\epsilon_{\rm xz}=\epsilon_{\rm yz}=0. 
\end{equation} 
Note that since $\sigma_{\rm zz}=0$, the transverse normal strain $\epsilon_{\rm zz}$, although not zero identically, does not appear in the total deformation energy and hence not in the equations of mechanical equilibrium. Consequently, it amounts to neglecting the transverse normal strain. Thus, we have, in theory, a case of \emph{both plane stress and plane strain}. In this case, the deformation of the thin Hydra laminate is due entirely to the in-plane displacement $u(x,y)\hat{\bf x}$ and $v(x,y)\hat{\bf y}$ and the out-plane deflection (bending) $\omega(x,y)$ of the neutral surface. The components of the displacement vector and the non-zero components of the strain tensor are given, respectively, by
\begin{subequations}\label{eq:Plates-dispStrain} 
\begin{equation}\label{eq:Plates-disp} 
u_{\rm x} \approx u(x,y)-(z-z_{\rm x 0}) \partial_{\rm x} \omega, \quad
u_{\rm y} \approx v(x,y)-(z-z_{\rm y 0}) \partial_{\rm y} \omega, \quad 
u_{\rm z} \approx \omega(x,y),
\end{equation}
\begin{equation}\label{eq:Plates-strain} 
\epsilon_{\rm xx}=\partial_{\rm x} u-(z-z_{\rm x 0})\partial_{\rm x}^{2} \omega, \quad
\epsilon_{\rm yy}=\partial_{\rm y} v-(z-z_{\rm y 0})\partial_{\rm y}^{2} \omega, \quad
\epsilon_{\rm xy}=\frac{1}{2}\left(\partial_{\rm y} u+ \partial_{\rm x} v\right)-\frac{1}{2}(z-z_{\rm x 0})(z-z_{\rm y 0})\partial_{\rm xy}^{2} \omega.  
\end{equation} 
\end{subequations}
Note that in contrast with the traditional CLPT, we here assume more generally that the positions of the neutral surfaces of the thin Hydra-laminated plate are different for the deformation in the $x$ and $y$ directions, and we denote them as $z_{\rm x 0}$ and $z_{\rm y 0}$, respectively. Such form of the displacement field allows reduction of the 3-D problem to one of studying the deformation of the reference neutral surfaces. Once the neutral-surface displacements $(u, v, \omega)$ are known, the displacements of any arbitrary point $(x, y, z)$ in the 3-D continuum can be determined.   

For the thin orthotropic Hydra-laminated plates, the total deformation energy (including the active work term~\cite{Xu2021,Xu2022}) is given by~\cite{Reddy2003} 
\begin{align}\label{eq:Plates-Ftxyz}
{\cal F}_{\rm t} =\int \int \int dx dy dz\left[ \left(\frac{1}{2} \bar{Q}_{11} \epsilon_{\rm xx}^{2}+\frac{1}{2} \bar{Q}_{22} \epsilon_{\rm yy}^{2}+\bar{Q}_{12} \epsilon_{\rm xx} \epsilon_{\rm yy} 
+2 \bar{Q}_{16} \epsilon_{\rm xy} \epsilon_{\rm xx}+2 \bar{Q}_{26} \epsilon_{\rm xy} \epsilon_{\rm yy}+2 \bar{Q}_{66} \epsilon_{\rm xy}^{2}\right)-\zeta^{(1)} \epsilon_{\rm x x}-\zeta^{(2)} \epsilon_{\rm y y}\right],
\end{align} 
in which the elastic constants are given by
\begin{align}\label{eq:Plates-Qbarij}
&\bar{Q}_{11}=Q_{11} \cos ^{4} \theta+2\left(Q_{12}+2 Q_{66}\right) \sin ^{2} \theta \cos ^{2} \theta+Q_{22} \sin ^{4} \theta, \\ \nonumber
&\bar{Q}_{12}=\left(Q_{11}+Q_{22}-4 Q_{66}\right) \sin ^{2} \theta \cos ^{2} \theta+Q_{12}\left(\sin ^{4} \theta+\cos ^{4} \theta\right), \\ \nonumber
&\bar{Q}_{22}=Q_{11} \sin ^{4} \theta+2\left(Q_{12}+2 Q_{66}\right) \sin ^{2} \theta \cos ^{2} \theta+Q_{22} \cos ^{4} \theta, \\ \nonumber
&\bar{Q}_{16}=\left(Q_{11}-Q_{12}-2 Q_{66}\right) \sin \theta \cos ^{3} \theta+\left(Q_{12}-Q_{22}+2 Q_{66}\right) \sin ^{3} \theta \cos \theta, \\ \nonumber
&\bar{Q}_{26}=\left(Q_{11}-Q_{12}-2 Q_{66}\right) \sin ^{3} \theta \cos \theta+\left(Q_{12}-Q_{22}+2 Q_{66}\right) \sin \theta \cos ^{3} \theta, \\ \nonumber
&\bar{Q}_{66}=\left(Q_{11}+Q_{22}-2 Q_{12}-2 Q_{66}\right) \sin ^{2} \theta \cos ^{2} \theta+Q_{66}\left(\sin ^{4} \theta+\cos ^{4} \theta\right),
\end{align} 
with the plane-stress-reduced orthotropic elastic constants in principal coordinates given by
\begin{align}\label{eq:Plates-Qij}
Q_{11}=\frac{E_{1}}{1-\nu_{12} \nu_{21}}, \quad Q_{12}=\frac{\nu_{12} E_{2}}{1-\nu_{12} \nu_{21}}, \quad 
Q_{22}=\frac{E_{2}}{1-\nu_{12} \nu_{21}}, \quad Q_{66}=G_{12}.
\end{align} 
Here $\theta$ is the angle measured counterclockwise from the $x$-direction to the (principal) direction of the supracellular cable in each Hydra lamina layer. Therefore, in the ectoderm layer with $\theta=0$, we have $\bar{Q}_{11}=Q_{11}$, $\bar{Q}_{12}={Q}_{12}$, $\bar{Q}_{22}={Q}_{22}$, $\bar{Q}_{16}=\bar{Q}_{26}=0$, and $\bar{Q}_{66}={Q}_{66}$. 
In the endoderm layer with $\theta=\pi/2$, we have $\bar{Q}_{11}=Q_{22}$, $\bar{Q}_{12}={Q}_{12}$, $\bar{Q}_{22}={Q}_{11}$, $\bar{Q}_{16}=\bar{Q}_{26}=0$, and $\bar{Q}_{66}={Q}_{66}$. 

The plane stress components in the $k$-th layer can be obtained from $\sigma_{\alpha\beta}^{
\rm e}=\delta {\cal F}_{\rm e}/\delta \epsilon_{\alpha\beta}$ (with ${\cal F}_{\rm e}$ being the elastic deformation energy in the round bracket of Eq.~(\ref{eq:Plates-Ftxyz})):
\begin{equation}\label{eq:Plates-sigmaepsilon}
 \begin{bmatrix}
\sigma_{\rm xx}^{
\rm e}\\
\sigma_{\rm yy}^{
\rm e}\\
\sigma_{\rm xy}^{
\rm e}
\end{bmatrix}^{(k)}=
\begin{bmatrix}
\bar{Q}_{11} & \bar{Q}_{12} & \bar{Q}_{16}\\
\bar{Q}_{12} & \bar{Q}_{22} & \bar{Q}_{26} \\
\bar{Q}_{16} & \bar{Q}_{26} & \bar{Q}_{66} 
\end{bmatrix}^{(k)}
 \begin{bmatrix}
\epsilon_{\rm xx}\\
\epsilon_{\rm yy}\\
2\epsilon_{\rm xy}
\end{bmatrix}^{(k)}. 
\end{equation}
Note that stresses are also linear through the thickness of each layer; however, they will have different linear variation in different material layers when $\bar{Q}_{i j}$ change from layer to layer. 


For a Hydra-plate suspended by its center with $\omega(0,0)=0$, we take the trial solution of the form 
\begin{equation}\label{eq:Plates-Mono-sol}
u=\epsilon_{\rm x} x, \quad v=\epsilon_{\rm y} y, \quad \omega = \frac{1}{2}\left( c_{\rm x} x^2 + c_{\rm y} y^2 \right),
\end{equation} 
and in this case, the total free energy in Eq. (\ref{eq:Plates-Ftxyz}) reduces to
\begin{align} \label{eq:Plates-Ftxyz2}
\mathcal{F}_{t}=&\int d x d y dz \left\{\left(\frac{1}{2} \bar{Q}_{11}\epsilon_{\rm x}^2+\bar{Q}_{12}\epsilon_{\rm x}\epsilon_{\rm y}+\frac{1}{2} \bar{Q}_{22}\epsilon_{\rm y}^2\right)-\left[\bar{Q}_{11}(z-z_{\rm x0})\epsilon_{\rm x}c_{\rm x}+\bar{Q}_{12}(z-z_{\rm y0})\epsilon_{\rm x}c_{\rm y}\right. \right.  \\ \nonumber
&\left. +\bar{Q}_{12}(z-z_{\rm x0})\epsilon_{\rm y}c_{\rm x} +\bar{Q}_{22}(z-z_{\rm y0})\epsilon_{\rm y}c_{\rm y}\right]  +\left[\frac{1}{2} \bar{Q}_{11}(z-z_{\rm x0})^2 c_{\rm x}^2+\bar{Q}_{12}(z-z_{\rm x0})(z-z_{\rm y0})c_{\rm x}c_{\rm y}\right.\\ \nonumber
& \left. \left.   +\frac{1}{2} \bar{Q}_{22}(z-z_{\rm y0})^2c_{\rm y}^2\right]-\left[\zeta^{(1)} \epsilon_{\rm x}+\zeta^{(2)} \epsilon_{\rm y}-\zeta^{(1)} (z-z_{\rm x0})c_{\rm x}-\zeta^{(2)} (z-z_{\rm y0})c_{\rm y}\right] \right\},
\end{align}
which can be integrated over the thickness ($z$) as
\begin{align} \label{eq:Plates-Ftxy}
\mathcal{F}_{t}=\int d x d y&\left[\left(\frac{1}{2} Y_{\rm p,11}\epsilon_{\rm x}^2+Y_{\rm p,12}\epsilon_{\rm x}\epsilon_{\rm y}+\frac{1}{2} Y_{\rm p,22}\epsilon_{\rm y}^2\right)-\left(\chi_{\rm p, 11}\epsilon_{\rm x}c_{\rm x}+\chi_{\rm p, 12}\epsilon_{\rm x}c_{\rm y}+\chi_{\rm p, 21} \epsilon_{\rm y}c_{\rm x}+\chi_{\rm p, 22}\epsilon_{\rm y}c_{\rm y}\right)+\right. \\ \nonumber
&\left. \left(\frac{1}{2} D_{\rm p,11}c_{\rm x}^2+D_{\rm p,12}c_{\rm x}c_{\rm y}+\frac{1}{2} D_{\rm p,22}c_{\rm y}^2\right)
 -\left(\tau_{\rm p}^{(1)} \epsilon_{\rm x}+\tau_{\rm p}^{(2)} \epsilon_{\rm y}+M_{\rm p}^{(1)} c_{\rm x}+M_{\rm p}^{(2)} c_{\rm y}\right)\right]. 
\end{align} 
Here $Y_{{\rm p},ij}$ are the extensional stiffnesses, $\chi_{{\rm p},ij}$ the bending-extensional coupling stiffnesses, and $D_{{\rm p},ij}$ are the bending stiffnesses, 
which are defined in terms of the lamina stiffnesses of each Hydra layer by 
\begin{align} \label{eq:Plates-stiffness}
& Y_{{\rm p},ij}\equiv \int_{-h_1}^{h_2} \bar{Q}_{i j} d z, \quad 
\left\{D_{{\rm p},11}, \, D_{{\rm p},12}, \,D_{{\rm p},22} \right\} \equiv\int_{-h_1}^{h_2} \left\{\bar{Q}_{11} (z-z_{\rm x 0})^{2}, \, \bar{Q}_{12} (z-z_{\rm x 0})(z-z_{\rm y 0}), \, \bar{Q}_{22} (z-z_{\rm y 0})^{2} \right\} d z\\ \nonumber 
& \left\{\chi_{{\rm p},11}, \, \chi_{{\rm p},12}, \, \chi_{{\rm p},21}, \,\chi_{{\rm p},22} \right\} \equiv\int_{-h_1}^{h_2} \left\{\bar{Q}_{11} (z-z_{\rm x 0}), \, \bar{Q}_{12}(z-z_{\rm y 0}), \, \bar{Q}_{12} (z-z_{\rm x 0}), \, \bar{Q}_{22} (z-z_{\rm y 0}) \right\} d z.
\end{align}
$\tau_{\rm p}^{(i)}<0$ are the active contractile forces per length and $M_{\rm p}^{(i)}$ are the active torques per length (generated by the asymmetric contraction), which are given in each layer, respectively, by
\begin{equation}\label{eq:Plates-tau12M12}
\tau_{\rm p}^{(1)}=\alpha_1\zeta_1 h_1, \quad
\tau_{\rm p}^{(2)}=\alpha_2\zeta_2 h_2, \quad
M_{\rm p}^{(1)}=\frac{1}{2} \tau_{\rm p}^{(1)}(2z_{\rm x 0}+\alpha_1h_{1})\approx \tau_{\rm p}^{(1)}z_{\rm x 0}, \quad
M_{\rm p}^{(2)}=\frac{1}{2} \tau_{\rm p}^{(2)}\left[2(z_{\rm y 0}-\alpha_{\rm m} h_2)-\alpha_2h_2\right]\approx \tau_{\rm p}^{(2)}z_{\rm y 0}.
\end{equation} 

In addition, we propose that, for a general thin composite active laminated plate (ALP), the position of the neutral surface can be obtained by considering a pure bending deformation with $u=v=0$ and $\omega\neq 0$, and by minimizing the corresponding bending energy  
\begin{align} \label{eq:Plates-Ftxyz2-bending}
\mathcal{F}_{t}=&\int d x d y dz \left[\frac{1}{2} \bar{Q}_{11}(z-z_{\rm x0})^2 c_{\rm x}^2+\bar{Q}_{12}(z-z_{\rm x0})(z-z_{\rm y0})c_{\rm x}c_{\rm y} +\frac{1}{2} \bar{Q}_{22}(z-z_{\rm y0})^2c_{\rm y}^2\right]. 
\end{align}
with respect to $z_{\rm x0}$ and $z_{\rm y0}$, giving 
\begin{equation}\label{eq:zx0zy0}
\chi_{{\rm p},ij}=0.    
\end{equation}
Obviously, this condition is over-determined: two unknown parameters $z_{\rm x0}$ and $z_{\rm y0}$ but with four independent equations. This is because, in general, the position of the neutral surface with zero strain is different for different strain components and depends in a more complex manner on the components of the pure-bending curvature tensor. In the following two subsections, we consider two special limiting cases where the above conditions are well-defined and the position of the neutral surface is uniquely-determined. 

\subsubsection{Limit of linear isotropic active-laminated plates}\label{sec:c0-Plates-Isotropic}

We first consider the limit of isotropic active-laminated plates where each lamina layer is linear isotropic and we have in each lamina layer: $E=E_1=E_2$, $\nu=\nu_{12}=\nu_{21}$, $\bar{Q}_{11}=\bar{Q}_{22}\equiv E_*\equiv E/(1-\nu^2)$, $\bar{Q}_{12}=\nu E_*$, and $\bar{Q}_{66}=E/2(1+\nu)$. To be specific, we only consider materials with $\nu>0$. In this limit, we first calculate the position of neutral surface. The free energy in Eq. (\ref{eq:Plates-Ftxyz2-bending}) becomes 
\begin{align}\label{eq:Plates-Ftxyz2-bending-isotropic}
\mathcal{F}_{t}=&\int d x d y dz \left\{\frac{1}{2}E_*\left[ (z-z_{\rm x0})^2 c_{\rm x}^2+2\nu(z-z_{\rm x0})(z-z_{\rm y0})c_{\rm x}c_{\rm y} +(z-z_{\rm y0})^2c_{\rm y}^2\right]\right\}. 
\end{align}
For simplicity, we further assume that the variation of the Poisson ratios $\nu$ through the thickness is negligible, and then the condition in Eq.~(\ref{eq:zx0zy0}) gives the position of neutral surface: $z_{\rm x0}=z_{\rm y0}=z_0$ with $z_0$ given by 
\begin{align}\label{eq:Plates-zx0zy0-isotropic}
\int dz \left[ E_*(z) (z-z_0)\right]=0, \quad {\rm or}, \quad
z_0=\frac{\int dz E_*(z) z}{\int dz E_*(z)}=\frac{E_{\rm{p,eff}}^{(2)} h_{2}^{2}-E_*^{(1)} h_{1}^{2}}{2\left(Y_{\rm p}^{(1)}+Y_{\rm{p, eff}}^{(2)}\right)},
\end{align}
with the compression modulus in the lamina layer-1 and the composite lamina layer-2 given by
\begin{equation}\label{eq:Plates-Y1Yeff2}
Y_{\rm p}^{(1)}= E_*^{(1)}h_1, \quad 
Y_{\rm{p, eff}}^{(2)}=E_*^{(\rm m)}h_{\rm m}+E_*^{(2)} (h_2-h_{\rm m}), \quad
\end{equation}
respectively, and $E_{\rm{p, eff}}^{(2)}\equiv {\int_0^{h_2} dz E_*(z) z}/{\int_0^{h_2} dz z}$. 
In addition, the total free energy in Eq. (\ref{eq:Plates-Ftxy}) then reduces to
\begin{align} \label{eq:Plates-Ftxy-isotropic}
\mathcal{F}_{t}=\int d x d y&\left[\frac{1}{2} Y_{\rm p}\left(\epsilon_{\rm x}^2+2\nu\epsilon_{\rm x}\epsilon_{\rm y}+\epsilon_{\rm y}^2\right)+ \frac{1}{2} D_{\rm p}\left(c_{\rm x}^2+2\nu c_{\rm x}c_{\rm y}+ c_{\rm y}^2\right)
 -\left(\tau_{\rm p}^{(1)} \epsilon_{\rm x}+\tau_{\rm p}^{(2)} \epsilon_{\rm y}+M_{\rm p}^{(1)} c_{\rm x}+M_{\rm p}^{(2)} c_{\rm y}\right)\right], 
\end{align} 
in which the effective compression modulus, $Y_{{\rm p}}$, and the effective flexural stiffness or rigidity, $D_{{\rm p}}$, are given by 
\begin{align} \label{eq:Plates-stiffness-isotropic}
& Y_{{\rm p}}\equiv \int_{-h_1}^{h_2} dz E_*(z)=Y_{\rm p}^{(1)}+Y_{\rm{p, eff}}^{(2)}, \quad
D_{{\rm p}} \equiv\int_{-h_1}^{h_2} dz E_*(z) (z-z_0)^{2}=D_{\rm p}^{(1)}+D_{\rm{p, eff}}^{(2)}.
\end{align}
and the active forces $\tau_{\rm p}^{(k)}$ and the active torques $M_{\rm p}^{(k)}$ are given in Eq. (\ref{eq:Plates-tau12M12}). 
Here the flexural stiffness in the lamina layer-1 and the composite lamina layer-2 are given, respectively, by
\begin{equation}\label{eq:Plates-Bilayer-EffectiveLayer21}
D_{\rm p}^{(1)}=E_*^{(1)}\left(\frac{h_1^3}{3}-z_0^2h_1\right), \quad 
D_{\rm{p, eff}}^{(2)}= E_{\rm{p, eff}}^{'(2)}\left(\frac{h_2^3}{3}-\frac{Y_{\rm{p, eff}}^{(2)}}{E_{\rm{p, eff}}^{'(2)}h_2}z_0^2h_2\right), 
\end{equation}
with $E_{\rm{p, eff}}^{'(2)}\equiv {\int_0^{h_2} dz E_*(z) z^2}/{\int_0^{h_2} dz z^2}$.
Minimization of ${\cal F}_{\rm t}$ in Eq. (\ref{eq:Plates-Ftxy-isotropic}) then gives the spontaneous contraction and curvature:
\begin{subequations}\label{eq:Plates-e0c0} 
\begin{equation}\label{eq:Plates-ec} 
\epsilon_{\rm x 0}=\frac{\epsilon_{1}-\nu \epsilon_{2}}{1-\nu^{2}} , \quad 
\epsilon_{\rm y 0}=\frac{\epsilon_{2}-\nu \epsilon_{1}}{1-\nu^{2}}, \quad 
c_{\rm x 0}=\frac{c_{1}-\nu c_{2}}{1-\nu^{2}}, \quad
c_{\rm y 0}=\frac{c_{2}-\nu c_{1}}{1-\nu^{2}},
\end{equation}
with
\begin{equation}\label{eq:Plates-e12c12} 
\epsilon_1=\frac{\tau_{\rm p}^{(1)}}{Y_{\rm{p}}}, \quad \epsilon_2=\frac{\tau_{\rm p}^{(2)}}{Y_{\rm{p}}}, \quad
c_1=\frac{M_{\rm p}^{(1)}}{D_{\rm{p}}}, \quad
c_2=\frac{M_{\rm p}^{(2)}}{D_{\rm{p}}}. 
\end{equation}
\end{subequations}

Note that Eqs. (\ref{eq:Plates-e0c0}) indicates that the spontaneous curvature or the equilibrium bent shape of Hydra tissue plates is determined by the anisotropy in the contractility of the two perpendicular supracellular cables along the two principal directions. The contractility anisotropy can be measured by the ratio $c_2/c_1$ and according to it the following two general shapes are determined. 
\begin{itemize}
    \item {\bf Elliptical cap shape with $c_{\rm x 0}c_{\rm y 0}>0$}. --- From Eqs. (\ref{eq:Plates-e0c0}), we get $\nu< c_2/c_1<\nu^{-1}$, no matter $c_{1,2}$ is positive or negative. 
    \item {\bf Saddle shape with $c_{\rm x 0}c_{\rm y 0}<0$}. --- From Eqs. (\ref{eq:Plates-e0c0}), we get $c_2/c_1>\nu^{-1}$ or $c_2/c_1<\nu$, no matter $c_{1,2}$ is positive or negative. 
\end{itemize} 
Plate-like Hydra fragments are observed in experiments~\cite{Livshits2017} to bend into the cap shape toward the endoderm-mesoglea layer (lamina layer-2). This corresponds to the case in the ALP model of $c_{\rm x0} \approx c_{\rm y0}>0$ and $c_1 \approx c_2>0$. That is, one the one hand, the contractility of the two perpendicular supracellular cables are more or less the same: $\tau_{\rm p}^{(1)} \approx \tau_{\rm p}^{(2)}$ and $M_{\rm p}^{(1)}\approx M_{\rm p}^{(2)}$. On the other hand, from Eqs. (\ref{eq:Plates-tau12M12}), (\ref{eq:Plates-zx0zy0-isotropic}), and for thin contracting layers (with $\alpha_1,\, \alpha_2 \to 0$), we find that such inward spontaneous bending of Hydra fragments (with $c_{\rm x0},\, c_{\rm y0}>0$) requires $z_{0}<0$, or equivalently,
\begin{equation}\label{eq:Plates-csign}
E_{\rm{p,eff}}^{(2)} h_{2}^{2}<E_*^{(1)} h_{1}^{2},
\end{equation} 
that is, the neutral surface is in the ectoderm layer-1. To reach this requirement, we would like to address the importance of the presence of a very soft mesoglea layer with $E_*^{(\rm m)} \ll E_*^{(2)}$, which can reduce the effective modulus 
$E_{\rm{p, eff}}^{(2)}=E_*^{(2)}(1- \alpha_{\rm m}^{2} + E_*^{(\rm m)}\alpha_{\rm m}^{2}/{E_*^{(2)}})$
of the composite endoderm-mesoglea layer (lamina layer-2) significantly down to $E_*^{(2)}(1- \alpha_{\rm m}^{2})\ll E_*^{(2)}$. Then from Eqs. (\ref{eq:Plates-csign}) and (\ref{eq:Plates-e12c12}), we see this softening can ensure $z_{0}<0$ and the condition $c_1, \, c_2>0$.

\subsubsection{Limit of strongly anisotropic active-laminated plates}\label{sec:c0-Plates-anisotropic}

We next consider the strongly anisotropic limit with $E_1^{(k)}\gg E_2^{(k)} \sim G_{12}^{(k)}$ and $\nu_{12}^{(k)} \sim 1$ in the $k$-th lamina layer with $k=1,{\rm m},2$. Then from $\nu_{12}^{(k)} E_2^{(k)}=\nu_{21}^{(k)} E_1^{(k)}$, we have $\nu_{21}^{(k)} \ll 1$. In this case, from Eqs.~(\ref{eq:Plates-Qbarij}) and (\ref{eq:Plates-Qij}) we have
\begin{subequations}\label{eq:strongly-anisotropic}
\begin{equation}
\bar{Q}_{11}^{(1)}\sim E_1^{(1)} \gg 1, \quad
\bar{Q}_{22}^{(1)}\sim E_2^{(1)} \sim 1,  \quad 
\bar{Q}_{12}^{(1)}\sim \nu_{12}^{(1)} E_2^{(1)} \sim 1,
\end{equation} 
\begin{equation}
\bar{Q}_{11}^{(2)}\sim E_2^{(2)} \sim 1, \quad
\bar{Q}_{22}^{(2)}\sim E_1^{(2)} \gg 1,  \quad 
\bar{Q}_{12}^{(2)}\sim \nu_{12}^{(2)} E_2^{(2)} \sim 1, 
\end{equation} 
\end{subequations}
and hence
\begin{equation}
\chi_{{\rm p},11}\sim \int_{-h_1}^{0} \bar{Q}_{11}^{(1)} (z-z_{\rm x 0}) dz, \, \chi_{{\rm p},12}\sim \chi_{{\rm p},21} \ll \chi_{{\rm p},11}, \,\chi_{{\rm p},22} \sim \int_{0}^{h_2} \bar{Q}_{22}^{(2)} (z-z_{\rm y 0}) dz. 
\end{equation}
Therefore, from the condition $\chi_{{\rm p},ij}=0$ in Eq. (\ref{eq:zx0zy0}) is identical to assuming $\chi_{{\rm p},11}=\chi_{{\rm p},22}=0$ and we obtain 
\begin{equation}
z_{\rm x 0}\approx -h_1/2, \quad z_{\rm y 0}=h_2/2.
\end{equation}
The total free energy in Eq. (\ref{eq:Plates-Ftxy}) can then be approximated by
\begin{align}\label{eq:Plates-Ftxy-anistropic}
\mathcal{F}_{t}\approx \int d x d y\left[\frac{1}{2} Y_{\rm p}^{(1)} \epsilon_{\rm x}^2+\frac{1}{2} Y_{\rm p}^{(2)}\epsilon_{\rm y}^2+
\frac{1}{2} D_{\rm p}^{(1)} c_{\rm x}^2+\frac{1}{2} D_{\rm p}^{(2)} c_{\rm y}^2 -\tau_{\rm p}^{(1)} \epsilon_{\rm x}-\tau_{\rm p}^{(2)} \epsilon_{\rm y}-M_{\rm p}^{(1)} c_{\rm x}-M_{\rm p}^{(2)} c_{\rm y}\right],
\end{align} 
with the effective compression moduli and flexural stiffnesses given by 
\begin{equation}
Y_{\rm p}^{(1)} \sim E_1^{(1)}h_1, \quad
D_{\rm p}^{(1)} \sim \frac{1}{3} E_1^{(1)}\left[(h_1+z_0)^3-z_0^3\right], \quad 
Y_{\rm p}^{(2)}\sim E_1^{(2)}h_2, \quad
D_{\rm p}^{(2)} \sim \frac{1}{3} E_1^{(2)}\left[(h_2-z_0)^3-(h_{\rm m}-z_0)^3\right]. 
\end{equation}  
Minimization of ${\cal F}_{\rm t}$ then gives the spontaneous contraction and curvature:
\begin{equation}\label{eq:Plates-anisotropic-ec} 
\epsilon_{\rm x 0}=\frac{\tau_{\rm p}^{(1)}}{Y_{\rm p}^{(1)}}, \quad \epsilon_{\rm y 0}=\frac{\tau_{\rm p}^{(2)}}{Y_{\rm p}^{(2)}}, \quad
c_{\rm x 0}=\frac{M_{\rm p}^{(1)}}{D_{\rm p}^{(1)}}, \quad 
c_{\rm y 0}=\frac{M_{\rm p}^{(2)}}{D_{\rm p}^{(2)}}. 
\end{equation}
That is, in the limit of strong anisotropic lamina, the contractions and bending induced by the two perpendicular supracellular actomyosin cables become independent. Furthermore, when the contracting layers and the mesoglea layer are very thin ($\alpha_1,\, \alpha_{\rm m}, \, \alpha_2 \to 0$), from Eq. (\ref{eq:Plates-tau12M12}) we have $M_{\rm p}^{(1)}\approx \tau_{\rm p}^{(1)}z_{\rm x 0} \approx -\tau_{\rm p}^{(1)}h_1/2>0$ and $M_{\rm p}^{(2)}\approx \tau_{\rm p}^{(2)}z_{\rm y 0}\approx \tau_{\rm p}^{(2)} h_2/2<0$. That is, the fragment plate is bent spontaneously to be a saddle shape. 

As a summary, from the calculations in Sec.~\ref{sec:c0-Plates-Isotropic} and Sec.~\ref{sec:c0-Plates-anisotropic}, we conclude that the spontaneous curvature or the equilibrium bent shape of Hydra tissue plates is determined by the anisotropy in both the supracellular actomyosin contractility (measured by the ratio $c_2/c_1$) and elastic properties of each lamina layer (measured by the ratio $E_2^{(k)}/E_1^{(k)}$ in the $k$-th layer). These results have been summarized in Fig. 2 of the main text.

\subsection{Rod-like Hydra tissue fragments}

When one side dimension is much smaller than the other in a Hydra tissue fragment plate, the plate becomes a rod. If, furthermore, one of the two lateral principal dimensions is much smaller than the other, the fragment is thought to be rod-like (as shown in Fig. 2(a) in the main text).  
In this subsection, we consider the spontaneous curvature of such rod-like Hydra fragments with a narrow rectangular cross-section. 

The fragment rod is thought to be thin if its longitudinal (along the direction of Hydra-body axis) contour length $\ell_{\rm c}$ of is much larger than its thickness $h$ and its lateral width $w$. In this case, we have the zero-normal stress condition~ \cite{Landau1986}: $\bm{\sigma}\cdot \hat{\bm n}=0$ with $\hat{\bm n}$ being the outward unit normal vector of the side surfaces. Therefore, all the components of the stress tensor except $\sigma_{\rm xx}$ must be zero, and $\epsilon_{\rm xy}=\epsilon_{\rm xz}=\epsilon_{y z}=0$ (however, $\epsilon_{\rm yy}, \, \epsilon_{\rm zz}\neq 0$). 
Note that since $\sigma_{{\rm zz}}=\sigma_{{\rm yy}}=0$, the transverse normal strains $\epsilon_{\rm zz}$ and $\epsilon_{\rm yy}$, although not zero identically, do not appear in the total free energy and hence in the equations of mechanical equilibrium. Consequently, it amounts to neglecting the transverse normal strains. In this case, the deformation of thin Hydra laminate rods is due entirely to the in-plane displacement $u(x)$ along longitudinal $x-$direction and the out-plane deflection (bending) $\omega(x)$ of the neutral line located at $z = z_0$. The longitudinal displacement and strain are given by
\begin{align}\label{eq:Rods-strain}
u_{\rm x} \approx u-(z-z_0) \omega', \quad  
\epsilon_{\rm xx}=u'-(z-z_0) \omega'',
\end{align}
respectively, with the prime hereafter denoting the ordinary derivatives. For the thin orthotropic Hydra-laminate rods, from Eq.~(\ref{eq:Plates-sigmaepsilon}) and vanishing $\sigma_{\rm yy}=\sigma_{\rm yy}^{\rm e}=0$, we have $\epsilon_{\rm yy}=-\bar{Q}_{12} \epsilon_{\rm xx}/\bar{Q}_{22}$. In this case, the total deformation energy for plates in Eq.~(\ref{eq:Plates-Ftxyz}) reduces to that of rods as 
\begin{align}\label{eq:Rods-Ftxyz}
{\cal F}_{\rm t} =\int_{-\ell_{\rm c}/ 2}^{\ell_{\rm c}/2} \int_{-w/2}^{w/2} \int dx dy dz\left(\frac{1}{2} \frac{\bar{Q}_{11}\bar{Q}_{22}-\bar{Q}_{12}^2}{\bar{Q}_{22}} \epsilon_{\rm xx}^{2}-\zeta^{(1)} \epsilon_{\rm x x}\right).
\end{align}
 
Moreover for rod-like Hydra tissue fragments, the active contraction in the endoderm layer can be neglected and we only consider the active contraction of supracellular actomyosin cables in the ectoderm layer (\emph{i.e.}, the layer-1 shown in Fig.~\ref{Fig:Schematic-Plates}(b)); its thickness and active stress are respectively given by $h_{\rm a}^{(1)}\equiv \alpha_1 h_1$ and ${\bm \sigma}^{\rm a(1)}=\zeta^{(1)} \hat{\bf x}\hat{\bf x}$ as in Eq.~(\ref{eq:Plates-hsigmaa}). 
Here $\alpha_1 \in [0,1]$ measures the fraction of contracting layer: $\alpha=0$ representing a purely passive rod, and $\alpha=1$ representing a symmetric active rod with uniform contractile stresses through the layer-1 thickness. 
$\zeta^{(1)}(z)<0$ for active contractile stresses in layer-1 follows a simple step-wise form as in Eq.~(\ref{eq:Plates-zeta}) (see Fig.~\ref{Fig:Schematic-Plates}(b)). 

For a Hydra-plate suspended by its center with $\omega(0,0)=0$, we take the trial solution of the form 
\begin{equation}\label{eq:Rods-Mono-sol}
u=\epsilon x, \quad \omega = \frac{1}{2} c x^2,
\end{equation}  
and in this case, the total free energy in Eq.~(\ref{eq:Rods-Ftxyz}) reduces to
\begin{align} \label{eq:Rods-Ftxyz2}
{\cal F}_{\rm t} =\int_{-\ell_{\rm c}/ 2}^{\ell_{\rm c}/2} \int_{-w/2}^{w/2} \int_{-h_1}^{h_2} dx dy dz\left\{\frac{1}{2}  \frac{\bar{Q}_{11}\bar{Q}_{22}-\bar{Q}_{12}^2}{\bar{Q}_{22}}\left[\epsilon^{2}- 2(z-z_0)\epsilon c +
(z-z_0)^2 c^{2}\right] -\zeta^{(1)} \epsilon+\zeta^{(1)} (z-z_0)c\right\},
\end{align}
which can be integrated over $z$ and $y$ as
\begin{equation}\label{eq:Rods-Ftx}
{\cal F}_{\rm t}(\epsilon,c) = \int_{-\ell_{\rm c}/ 2}^{\ell_{\rm c}/2} d x\left(\frac{1}{2} Y_{\rm{r}}\epsilon^{2}-\chi_{{\rm r}}\epsilon c +\frac{1}{2} D_{\rm{r}}c^{2}-\tau_{\rm{r}} \epsilon-M_{\rm{r}} c\right).
\end{equation}
Here $Y_{{\rm r}}$ are the extensional stiffnesses, $\chi_{{\rm r}}$ the bending-extensional coupling stiffnesses, and $D_{{\rm r}}$ are the bending stiffnesses, 
which are defined in terms of the lamina stiffnesses of each Hydra layer by 
\begin{align} \label{eq:Plates-stiffness}
& Y_{{\rm r}}\equiv \int_{-h_1}^{h_2} \frac{\bar{Q}_{11}\bar{Q}_{22}-\bar{Q}_{12}^2}{\bar{Q}_{22}} d z, \quad 
D_{{\rm r}} \equiv\int_{-h_1}^{h_2} \frac{\bar{Q}_{11}\bar{Q}_{22}-\bar{Q}_{12}^2}{\bar{Q}_{22}} (z-z_{\rm x 0})^{2} d z, \quad 
\chi_{{\rm r}} \equiv\int_{-h_1}^{h_2} \frac{\bar{Q}_{11}\bar{Q}_{22}-\bar{Q}_{12}^2}{\bar{Q}_{22}} (z-z_0) d z.
\end{align}
$\tau_{{\rm r}}<0$ are the active contractile forces per length and $M_{\rm r}$ are the active torques per length (generated by the asymmetric contraction), given by
\begin{equation}\label{eq:Rods-tau12M12}
\tau_{{\rm r}}=\alpha_1\zeta_1 h_1, \quad
M_{\rm r}=\frac{1}{2} \tau_{\rm r}(2z_0+\alpha_1h_{1})\approx \tau_{\rm r}z_0.
\end{equation} 
Note that the general expressions for the neutral line position $z_0$, the compression modulus $Y_{\rm{r}}$, and the flexural stifness $D_{\rm{r}}$ are complex; we consider only two particularly interesting limits in the following discussion. 

In addition, we propose that, for a general thin composite laminated rod, the position of the neutral line can be obtained by considering a pure bending deformation with $u=0$ and $\omega\neq 0$, and by minimizing the corresponding bending energy  
\begin{align} \label{eq:Rods-Ftxyz2-bending}
\mathcal{F}_{t}=
\int dx dy dz\left[\frac{1}{2}  \frac{\bar{Q}_{11}\bar{Q}_{22}-\bar{Q}_{12}^2}{\bar{Q}_{22}}
(z-z_0)^2 c^{2}\right].
\end{align}
with respect to $z_0$ giving 
\begin{equation}\label{eq:Rods-z0}
\chi_{{\rm r}}=0, \quad {\rm or}, \quad
z_0= \int_{-h_1}^{h_2} \frac{\bar{Q}_{11}\bar{Q}_{22}-\bar{Q}_{12}^2}{\bar{Q}_{22}} z d z\bigg/\int_{-h_1}^{h_2} \frac{\bar{Q}_{11}\bar{Q}_{22}-\bar{Q}_{12}^2}{\bar{Q}_{22}} d z. 
\end{equation}

\subsubsection{Limit of linear isotropic active-laminated rods}

As mentioned above, in the isotropic limit, we have in each lamina layer: $E=E_1=E_2$, $\nu=\nu_{12}=\nu_{21}$, $\bar{Q}_{11}=\bar{Q}_{22}\equiv E_*=E/(1-\nu^2)$, $\bar{Q}_{12}=\nu E_*$, and $\bar{Q}_{66}=E/2(1+\nu)$. Therefore, we have $E=({\bar{Q}_{11}\bar{Q}_{22}-\bar{Q}_{12}^2})/{\bar{Q}_{22}}$ in each layer and the neutral line position $z_0$ can be calculated from Eq.~(\ref{eq:Rods-z0}) to be  
\begin{equation}\label{eq:Rods-Trilayer-z0}
z_{0}=\frac{E_{\rm{r,eff}}^{(2)}w h_{2}^{2}-E_{1}w h_{1}^{2}}{2\left(Y_{\rm r}^{(1)}+Y_{\rm{r, eff}}^{(2)}\right)},
\end{equation} 
with the compression modulus in the lamina layer-1 and the composite lamina layer-2 given by
\begin{equation}\label{eq:Rods-Y1Yeff2}
Y_{\rm r}^{(1)}= E^{(1)}w h_1, \quad 
Y_{\rm{r, eff}}^{(2)}=E^{(\rm m)} w h_{\rm m}+E^{(2)} w (h_2-h_{\rm m}), \quad
\end{equation}
respectively, and $E_{\rm{r, eff}}^{(2)}\equiv {\int_0^{h_2} dz E(z) z}/{\int_0^{h_2} dz z}$. 
Note that the neutral line lies in the lamina layer-1 (\emph{i.e.}, $z_{0}<0$), when $E_{\rm{r,eff}}^{(2)} h_{2}^{2}<E^{(1)} h_{1}^{2}$.
Furthermore in this limit, the total free energy in Eq. (\ref{eq:Rods-Ftx}) reduces to 
\begin{equation}\label{eq:Rods-Ftx-isotropic}
{\cal F}_{\rm t}(\epsilon,c) = \int_{-\ell_{\rm c}/ 2}^{\ell_{\rm c}/2} d x\left(\frac{1}{2} Y_{\rm{r}}\epsilon^{2}+\frac{1}{2} D_{\rm{r}}c^{2}-\tau_{\rm{r}} \epsilon-M_{\rm{r}} c\right),
\end{equation} 
in which the effective compression modulus, $Y_{{\rm r}}$, and the effective flexural stiffness or rigidity, $D_{{\rm r}}$, are given by 
\begin{align} \label{eq:Rods-stiffness-isotropic}
& Y_{{\rm r}}\equiv \int_{-h_1}^{h_2} dz E(z)=Y_{\rm r}^{(1)}+Y_{\rm{r, eff}}^{(2)}, \quad
D_{{\rm r}} \equiv\int_{-h_1}^{h_2} dz E(z) (z-z_0)^{2}=D_{\rm r}^{(1)}+D_{\rm{r, eff}}^{(2)}. 
\end{align}
and the active forces $\tau_{\rm r}$ and the active torques $M_{\rm r}$ are given in Eq. (\ref{eq:Rods-tau12M12}). 
Here the flexural stiffness in the lamina layer-1 and the composite lamina layer-2 are given, respectively, by
\begin{equation}\label{eq:Rods-D1Deff2} 
D_{\rm r}^{(1)}=E^{(1)}w\left(\frac{h_1^3}{3}-z_0^2h_1\right), \quad 
D_{\rm{r, eff}}^{(2)}= E_{\rm{p, eff}}^{'(2)}w\left(\frac{h_2^3}{3}-\frac{Y_{\rm{r, eff}}^{(2)}}{E_{\rm{r, eff}}^{'(2)}h_2}z_0^2h_2\right), 
\end{equation}
with $E_{\rm{r, eff}}^{'(2)}\equiv {\int_0^{h_2} dz E(z) z^2}/{\int_0^{h_2} dz z^2}$. Minimization of ${\cal F}_{\rm t}$ with respect to $\epsilon$ and $c$ then again gives the spontaneous contraction and curvature:
\begin{equation}\label{eq:Rods-Trilayer-ec}
\epsilon_0=\frac{\tau_{\rm r}}{Y_{\rm{r}}}<0, \quad
c_0=\frac{M_{\rm r}}{D_{\rm{r}}}, 
\end{equation}
from which we obtain the condition (by assuming the contracting layer to be very thin, $\alpha_1\to 0$) for inward spontaneous bending (\emph{i.e.}, $c_0>0$): 
\begin{equation}\label{eq:Rods-Trilayer-csign}
E_{r,\rm{eff}}^{(2)}h_{2}^{2}<E^{(1)} h_{1}^{2}.
\end{equation} 
This condition also indicates that $z_{0}<0$ and the neutral line lies in the lamina layer-1. Similarly as discussed in Sec.~\ref{sec:c0-Plates-Isotropic}, the presence of a very soft mesoglea layer with $E^{(\rm m)} \ll E^{(2)}$ can reduce the effective modulus $E_{{\rm{r,eff}}}^{(2)}=E^{(2)}(1- \alpha_{\rm m}^{2} + \alpha _{\rm m} ^{2}E^{(\rm m)}/E^{(2)})$ of the composite endoderm-mesoglea layer (lamina layer-2) significantly down to $E^{(2)}(1-\alpha _{\rm m}^{2})\ll E^{(2)}$. This softening ensures the condition (\ref{eq:Rods-Trilayer-csign}) so that the supracellular actomyosin cable always bends the Hydra tissue fragment inward toward endoderm layer.


\subsubsection{Limit of strongly anisotropic active-laminated rods}

In the strongly anisotropic limit, we obtain from Eq. (\ref{eq:strongly-anisotropic}) that in the lamina layer-1 $({\bar{Q}_{11}\bar{Q}_{22}-\bar{Q}_{12}^2})/{\bar{Q}_{22}} \sim E_1^{(1)}$ which is much larger than that in the lamina layer-2 and hence the neutral line position $z_0$ can be calculated from Eq.~(\ref{eq:Rods-z0}) to be 
\begin{equation}
z_0\approx -h_1/2.
\end{equation}
The total free energy still takes the same form of Eq. (\ref{eq:Rods-Ftx-isotropic}) 
with the effective compression moduli and flexural stiffnesses given by 
\begin{equation}
Y_{{\rm r}} \sim E_1^{(1)}w h_1, \quad
D_{{\rm r}} \sim \frac{1}{3} E_1^{(1)}w \left[(h_1+z_0)^3-z_0^3\right]\approx \frac{1}{12}E_1^{(1)}w h_2^3. 
\end{equation}  
Minimization of ${\cal F}_{\rm t}$ then gives 
\begin{equation}\label{eq:Rods-anisotropic-ec} 
\epsilon_0=\frac{\tau_{\rm r}}{Y_{{\rm r}}}, \quad
c_0=\frac{M_{\rm{r}}}{D_{{\rm r}}}. 
\end{equation} 
Particularly, when the contracting layers and the mesoglea layer are very thin ($\alpha_1,\, \alpha_{\rm m}, \, \alpha_2 \to 0$), we have $M_{\rm r}\approx \tau_{\rm r}z_0 \approx -\tau_{\rm p}^{(1)}h_1/2>0$. 
That is, in the limit of strong anisotropic lamina, the rod-like Hydra tissue fragment behaves like a single-layer rod~\cite{Xu2022} that is being contracted and bent inward (toward the inner endoderm layer) by the thin actomyosin cables at its top surfaces.

In this work, our major purpose is to study the spontaneous bending of Hydra tissue fragments and therefore, we neglect the in-plane contraction $\epsilon$ and simply write the total energy $\mathcal{F}_{\rm t}$ as the bending energy ${\cal F}_{\rm b}$ in the following simple phenomenological form of
\begin{align}\label{eq:Rods-Ft0}
{\cal F}_{\rm t}={\cal F}_{\rm b}= \int_{-\ell_{\rm c}/2}^{\ell_{\rm c}/2} ds \left[\frac{D_{\rm r}}{2} \left(c-c_{0}\right)^{2}\right], 
\end{align}
in which case, the equilibrium state is characterized by the spontaneous curvature $c_0$. 


\subsection{Mapping to Lewis’ brass-bar-and-rubber-band model}

\begin{figure}[htbp]
  \centering
\includegraphics[width=0.8\columnwidth]{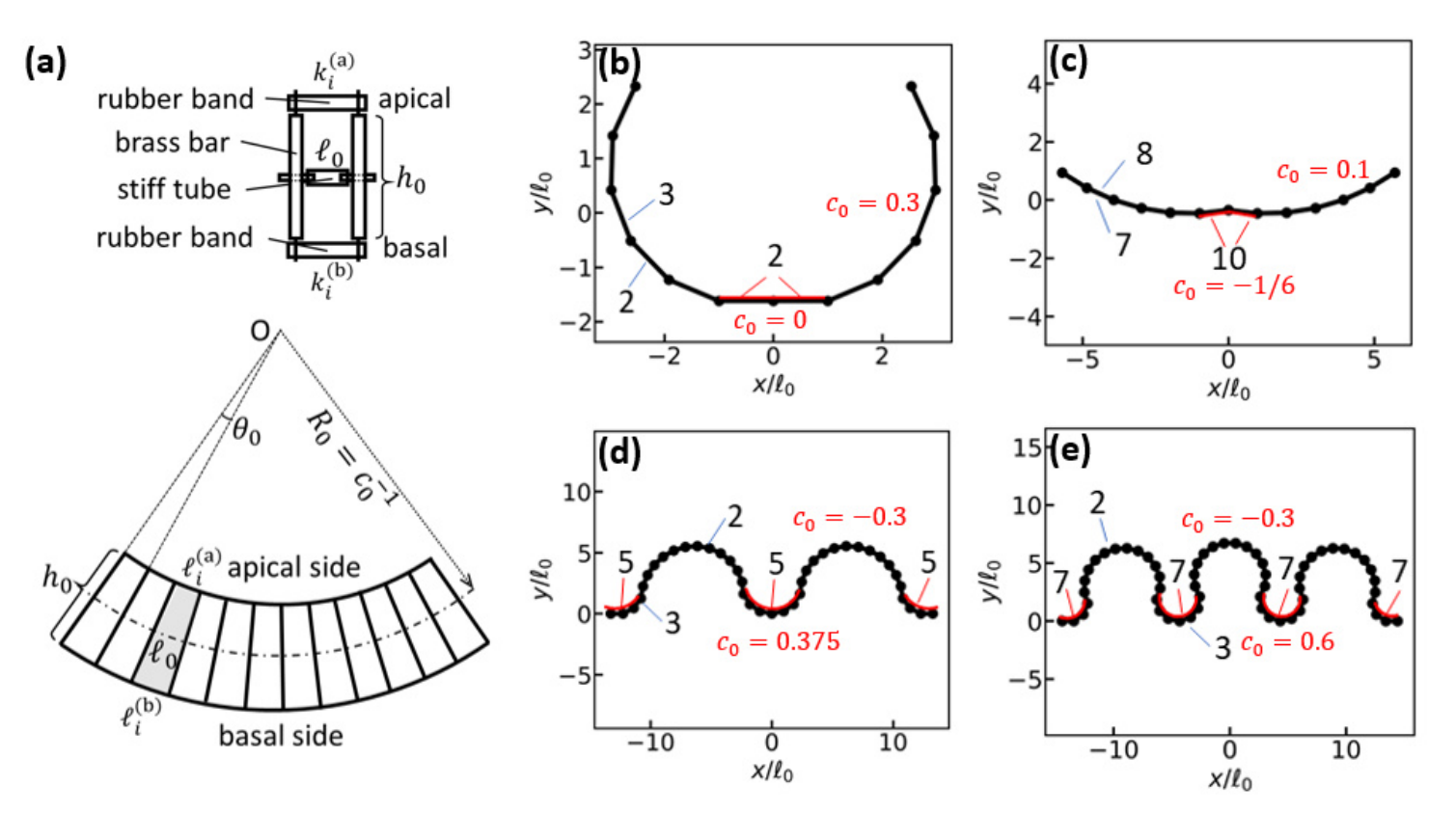}
  \caption {(a) Schematic illustrations of Lewis' mechanical model of epithelium consisting of brass bars, stiff rubber tubes, and rubber bands assembled into a series of quadrilateral segments each representing the cross-section of a cell. By adjusting the tension (or the effective spring constants $k_{i}^{\rm (a)}$ and $k_{i}^{\rm (b)}$) of the rubber bands on the basal and apical sides, the series can be deformed into arches of various curvatures $c_0$ (and the curvature radius $R_0=c_0^{-1}$). (b-e) A few nontrivial numerical solutions of the bead-spring model for Hydra rods with differential spontaneous curvature $c_0$ ($c_0$ are shown by red numbers). A simple mapping is found between Lewis' model and our ALP model of Hydra rods with non-zero spontaneous curvature. Note that the corresponding effective spring constants on the two sides in Lewis' model are shown by black numbers. (b) and (c) are the equilibrium shapes for unconstrained rods of contour length $\ell_{\rm c} = 12 \ell_0$ and (d-e) are for constrained rods of contour length $\ell_{\rm c} = 38 \ell_0$, $57 \ell_0$, respectively. 
  } \label{Fig:Schematic-Lewis}
\end{figure}

We would like to point out that the mechanistic perspective of tissue morphogenesis conveyed in this work is hardly new~\cite{Siber2017} and several classical theories have been developed. Among the earliest theories of this kind, W. H. Lewis proposed \cite{Lewis1947} (in 1947) a mechanical (experiment) model of epithelial sheets, consisting of brass bars, hinged at their centers to stiff rubber tubes (of a fixed length $\ell_0$) and tied, each to its neighbors, by stretched rubber bands at the two ends (see Fig.~\ref{Fig:Schematic-Lewis}). The brass bars, shared by the adjacent segments, represent the lateral interfaces of neighbouring cells. The stiff tubes mimic the incompressible cytoplasm by ensuring that the area of each segment is more or less fixed, while rubber bands model tension-generating machinery at the basal and apical surfaces of cells. 

Here we show for the spontaneous bending of Hydra rods that our mechanical ALP model has a simple mapping to Lewis' mechanical brass-bar-and-rubber-band model. In the Lewis' mechanical model, the total bending energy reads 
\begin{equation}\label{eq:Lewis-Fb}
{\cal F}_{\rm b}=\frac{1}{2} \sum_{i=1}^{N} \left[k_{i}^{\rm (a)} (\ell_{i}^{\rm (a)})^{2}+k_{i}^{\rm (b)}(\ell_{i}^{\rm (b)})^{2}\right], 
\end{equation} 
in which $k_{i}^{\rm (a)}$ and $k_{i}^{\rm (b)}$ are the effective spring constants, $\ell_{i}^{\rm (a)}$ and $\ell_{i}^{\rm (b)}$ are the lengths of the apical and basal sides, and $N$ denotes the total number of segments. The brass bars and stiff tubes together impose a geometric constraint requiring, 
\begin{equation}\label{eq:Lewis-lalb}
\ell_{i}^{\rm (a)}+\ell_{i}^{\rm (b)}=2 \ell_0, \quad {\rm or}, \quad
\ell_{i}^{\rm (b)}=2 \ell_0-\ell_{i}^{\rm (a)},
\end{equation}
with $\ell_0$ being the fixed length of the stiff tubes. Note that here we have assumed that the equilibrium lengths of the rubber bands are much smaller than $\ell_0$ so that they are taken simply to be zero and the final equilibrium configuration depends only on the spring constants. 

Particularly, for unconstrained sheets, the solution of Lewis' model is rather simple as the minimization of the total energy (\ref{eq:Lewis-Fb}), $\partial {\cal F}_{\rm b}/\partial {\ell_{i}^{\rm (a)}}=0$ (using Eq.~(\ref{eq:Lewis-lalb})), reduces to minimization of energies of each individual segment independently:
\begin{equation}\label{eq:Lewis-la0lb0}
\ell_{i}^{\rm (a)}=\frac{2k_{i}^{\rm (b)}\ell_0}{k_{i}^{\rm (a)}+k_{i}^{\rm (b)}}, \quad \ell_{i}^{\rm (b)}=\frac{2k_{i}^{\rm (a)}\ell_0}{k_{i}^{\rm (a)}+k_{i}^{\rm (b)}}. 
\end{equation} 
In addition, we note, from Fig.~\ref{Fig:Schematic-Lewis}, two geometric identities:
\begin{equation}\label{eq:Lewis-clalb}
\ell_{i}^{\rm (a)}=\ell_0(1-h_0c_{i}/2), \quad
\ell_{i}^{\rm (b)}=\ell_0(1+h_0c_{i}/2),
\end{equation}
and then using Eq.~(\ref{eq:Lewis-la0lb0}) we calculate the local equilibrium curvature as
\begin{equation}\label{eq:Lewis-c0}
c_{i,0}=\frac{2(k_{i}^{\rm (a)}-k_{i}^{\rm (b)})}{h_0(k_{i}^{\rm (a)}+k_{i}^{\rm (b)})},
\end{equation} 
with $h_0$ being the length of the brass bars. Using Eqs.~(\ref{eq:Lewis-clalb}) and (\ref{eq:Lewis-c0}), we can rewrite the bending energy in Eq.~(\ref{eq:Lewis-Fb}) as
\begin{equation}\label{eq:Lewis-Fb2}
{\cal F}_{\rm b}= \sum_{i=1}^{N}\frac{1}{2} D_{{\rm r}i} \ell_0 (c_i-c_{i,0})^2, 
\end{equation} 
in which $D_{{\rm r}i}\equiv (k_{i}^{\rm (a)}+k_{i}^{\rm (b)})h_0^2\ell_0/4$ is the flexural stiffness and a term ${2k_{i}^{\rm (a)}k_{i}^{\rm (b)}\ell_0^2}/({k_{i}^{\rm (a)}+k_{i}^{\rm (b)}})$ (that is constant for a given rod) has been neglected. The bending energy in Lew's model shown in Eq.~(\ref{eq:Lewis-Fb2}) is simply the discrete form of the bending energy in Eq.~(\ref{eq:Rods-Ft0}) in our composite Hydra ALP model. 
 
Note that the spontaneous curvature (or the local equilibrium curvature) $c_{i,0}$ is non-zero only if there exists some asymmetry as we have found above for composite Hydra tissue fragments. In Lewis' model, the spontaneous bending is driven simply by the asymmetry in effective spring constants, $(k_{i}^{\rm (a)}-k_{i}^{\rm (b)})$, on the apical and basal sides, while in the composite Hydra active-laminated rods, the spontaneous bending is driven by the active contraction of supra-cellular cables and is determined by the asymmetry in the mismatch between supra-cellular contractile cables and the position of the neutral line. Note that in both models the spontaneous bending is toward the softer side with smaller stiffnesses, for example, in Lewis' model, from Eq.~(\ref{eq:Lewis-c0}) we have $c_{i,0}>0$ if $k_{i}^{\rm (a)}>k_{i}^{\rm (b)}$, that is, the bending is toward the softer apical side.  

\subsection{Effects of inter-layer sliding \label{sec:sliding}} 



Up to now, we have only considered the case of coherent Hydra laminates where the Hydra lamina layers are perfectly bonded together. Here we consider the deformation of incoherent Hydra rods where sliding is assumed to be present only between between the ectoderm layer (\emph{i.e.}, the lamina layer-1) and the ectoderm-mesoglea composite layer (\emph{i.e.}, the lamina layer-2). 

We begin by considering thin Hydra rods that are contracted and bent uniformly where the spontaneous (out-plane) bending and in-plane contraction of the thin Hydra rods are driven by the uniformly contracting supracellular actomyosin cables in the lamina layer-1. The presence of inter-layer sliding indicates that the in-plane strain may differ (misfit or be discontinuous) from the lamina layer-1 to the lamina layer-2. In this case, the deformation of a thin rod can be determined by the in-plane displacement $u(x)$ along longitudinal $x-$direction, the out-plane deflection $\omega(x)$ of the neutral line at $z=z_0$, and the misfit strain (or the slide) $\epsilon_{\rm s}^{(k)}$ in lamina layer-$k$ relative to the neutral line. The longitudinal strain components are then given, respectively, by
\begin{align}\label{eq:Sliding-Rods-strain} 
\epsilon_{\rm xx}=\epsilon-(z-z_0) c + \epsilon_{\rm s}^{({\rm n})},
\end{align} 
where we simply take $\epsilon_{\rm s}^{({\rm 2})}=-\epsilon_{\rm s}^{({\rm 1})}$ and denote it by $\epsilon_{\rm s}$ (because we assume significant slide occurs only between the ectoderm layer and the mesoglea layer). Particularly, we only consider the linear isotropic limit, in which we have  $E=({\bar{Q}_{11}\bar{Q}_{22}-\bar{Q}_{12}^2})/{\bar{Q}_{22}}$ and the total energy is then given from Eq. (\ref{eq:Rods-Ftxyz}) by
\begin{align}\label{eq:Sliding-Rods-Ft} 
{\cal F}_{\rm t}=
\int dx \left[\frac{1}{2}Y_{\rm{r}}{\epsilon}^{2}+\frac{1}{2}Y_{\rm{s}}{\epsilon}_{\rm s}^{2}+\frac{1}{2} D_{\rm{r}} c^{2}+\Delta {Y}_{\rm r}\epsilon\epsilon_{\rm s}-\bar{Y}_{\rm r}hc\epsilon_{\rm s}-\tau_{\rm{r}} (\epsilon -{\epsilon}_{\rm s})-M_{\rm{r}} c\right],
\end{align}
in which $Y_{\rm r}=Y_{\rm r}^{(1)}+Y_{\rm r, eff}^{(2)}$, $\Delta {Y}_{\rm r} \equiv Y_{\rm r, eff}^{(2)}-Y_{\rm r}^{(1)}$, and $\bar{Y}_{\rm r} \equiv Y_{\rm r}^{(1)}Y_{\rm r, eff}^{(2)}/Y_{\rm r}$ with $Y_{\rm r}^{(1)}$ and $Y_{\rm r, eff}^{(2)}$ being the compression moduli in the lamina layer-1 and the composite lamina layer-2, respectively. Here we have introduced an additional energy contribution due to the inter-layer sliding, $\frac{1}{2}k_{\rm s}\epsilon_{\rm s}^2 $, per unit volume and $Y_{\rm s}\equiv {Y}_{\rm r}+k_{\rm s}wh$ with the elastic constant $k_{\rm s}$ characterizing the stiffness of inter-layer sliding. Minimizing $\mathcal{F}_{\rm t}$ with respect to $\epsilon$, $c$, and $\epsilon_{\rm s}$ gives the in-plane contraction, spontaneous curvature, and equilibrium inter-layer slide as 
\begin{equation}\label{eq:Sliding-Rods-c0es0} 
{\epsilon}_{\rm 0}= \frac{\tau_{\rm r}}{Y_{\rm r}}-\frac{\Delta Y_{\rm r}}{Y_{\rm r}}{\epsilon}_{\rm s0}, \quad 
c_0 = \frac{M_{\rm{r}}}{D_{\rm r}} + \frac{\bar{Y}_{\rm r}h}{D_{\rm r}}\epsilon_{\rm s0}, \quad 
{\epsilon}_{\rm s0}=\frac{M_{\rm{r}}h}{D_{\rm r}} \frac{\bar{Y}_{\rm r}}{\bar{Y}_{\rm s}} - \frac{\tau_{\rm r}}{\bar{Y}_{\rm s}}\left(1+\frac{\Delta Y_{\rm r}}{Y_{\rm r}}\right),
\end{equation} 
respectively, with $\bar{Y}_{\rm s}\equiv Y_{\rm s}-\bar{Y}_{\rm r}^2 h^2 /D_{\rm r}-{\Delta Y_{\rm r}}^2/{Y_{\rm r}}$. From Eq. (\ref{eq:Sliding-Rods-c0es0}), we can see that the coherent limit with $\epsilon_{\rm s} \to 0$ corresponds to Hydra rods with a very large inter-layer sliding stiffness, $Y_{\rm s}$, and hence ${Y}_{\rm s}, \, \bar{Y}_{\rm s}\gg {Y}_{\rm r}, \, \bar{Y}_{\rm r}$ and ${\tau_{\rm r}}/{\bar{Y}_{\rm s}} \to 0$. 


In this work, our major purpose is not to uncover the mechanisms for the inter-layer slide in Hydra tissue fragments, but to study the effects of inter-layer slide accompanying with spontaneous bending on their dynamics. Therefore, for a Hydra rod with differential slide and spontaneous curvature along the arc length, we neglect the in-plane contraction $\epsilon$ and simply write the total energy $\mathcal{F}_{\rm t}$ in the following simple phenomenological form (in comparison to that of coherent rods in Eq.~(\ref{eq:Rods-Ft0})) of
\begin{align}\label{eq:Sliding-Rods-Ft1} 
{\cal F}_{\rm t}= \int_{-\ell_{\rm c}/2}^{\ell_{\rm c}/2} ds \left[\frac{Y_{\rm s}}{2}\epsilon_{\rm s}^2 -\chi \epsilon_{\rm s} c+ \frac{\tilde{D}_{\rm r}}{2} \left(c-\tilde{c}_{0}\right)^{2}\right], 
\end{align}
with $Y_{\rm s}$ and $\tilde{D}_{\rm r}$ being the interlayer sliding stiffness and flexural stiffness of the rod, respectively. Here the first term is the energy related to the inter-layer sliding, the third term is the bending energy, and the second term is the contribution from the coupling between interlayer sliding and and the bending, respectively. Minimizing $\mathcal{F}_{\rm t}$ with respect to $c$ and $\epsilon_{\rm s}$ give the equilibrium conditions:
\begin{align}\label{eq:Sliding-Rods-Ft1-eq} 
c=c_0\equiv \frac{\tilde{c}_{0}}{1-\chi^2/\tilde{D}_{\rm r}Y_{\rm s}}, \quad \epsilon_{\rm s}= \epsilon_{\rm s0}\equiv \frac{\chi c_0}{Y_{\rm s}},
\end{align}
with $c_0$ and $\epsilon_{\rm s0}$ being the spontaneous curvature and the equilibrium slide, respectively. Using Eq.~(\ref{eq:Sliding-Rods-Ft1-eq}), we can further rewrite the total energy in Eq.~(\ref{eq:Sliding-Rods-Ft1}) into an alternative form as
\begin{align}\label{eq:Sliding-Rods-Ft0} 
\mathcal{F}_{\rm t}=\int_{-\ell_{\rm c}/2}^{\ell_{\rm c}/2} ds \left[\frac{Y_{\rm{s}}}{2}({\epsilon}_{\rm s}-{\epsilon}_{\rm s0}c/c_0)^{2}+\frac{D_{\rm{r}}}{2} (c-c_0)^{2}\right],
\end{align}
with $D_{\rm{r}}\equiv \tilde{D}_{\rm r}-\chi^2/Y_{\rm{s}}$ being the normalized flexural stiffness of the rod.

\section{Spontaneous bending dynamics of short Hydra rods \label{sec:BendDyn}}

Hydra tissue fragments cut from adult Hydra body bend spontaneously in their viscous fluid environment. In this section, we study the temporal evolution of the end-to-end distance of rod-like Hydra tissue fragments by combining approximate variational analysis based Onsager's variational principle with numerical simulations based on the bead-spring model.

\subsection{Analytical and simulation methods \label{sec:BendDyn-methods}} 

\subsubsection{Approximate variational method based on Onsager's variational principle \label{sec:BendDyn-methods-Onsager}} 

\begin{figure}[htbp]
  \centering
  \includegraphics[clip=true, viewport=1 1 800 400, keepaspectratio, width=0.6\textwidth]{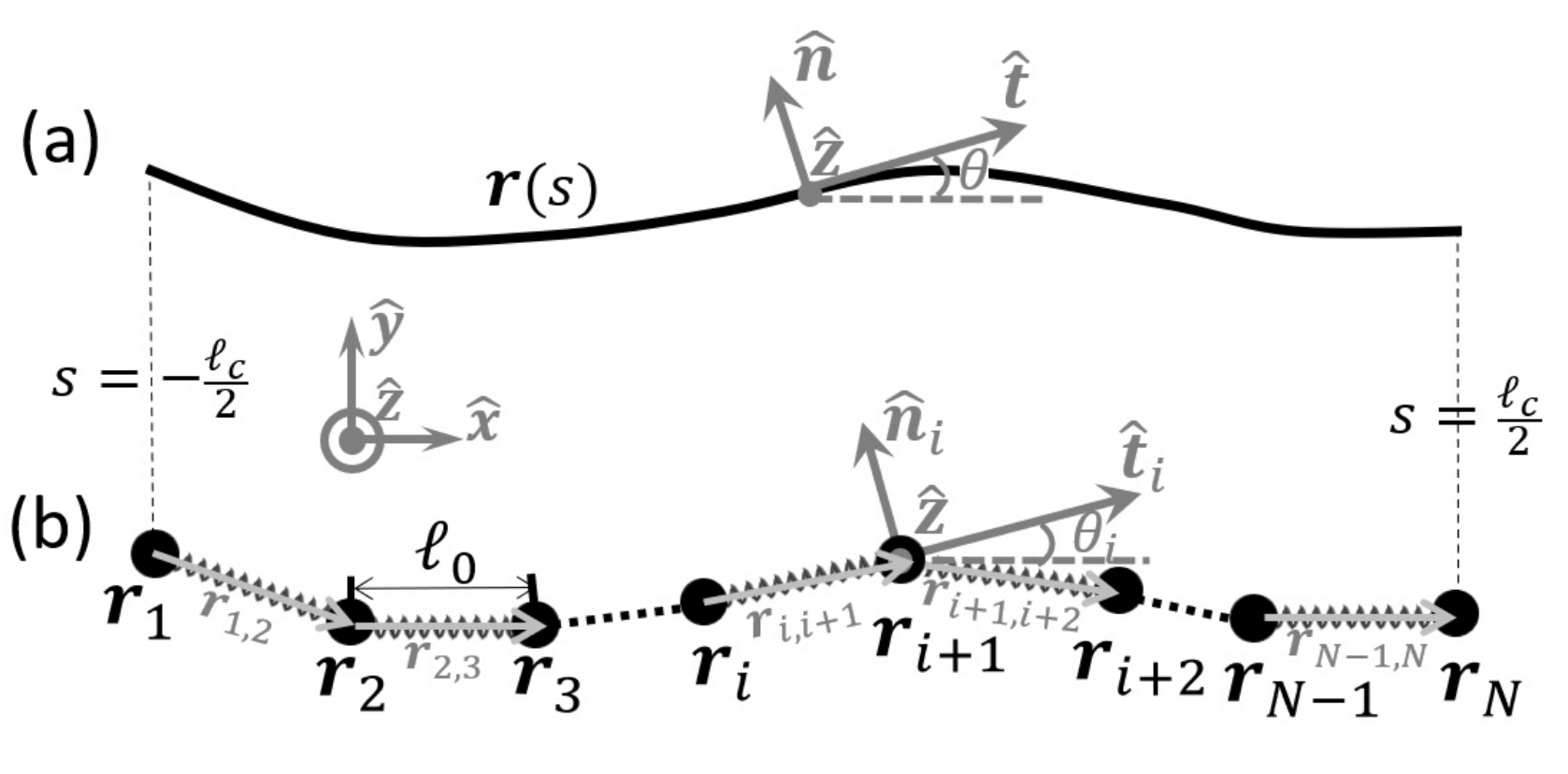}
  \caption {Schematic illustrations of the continuous elastica-curve model in (a) and the discrete bead-spring model in (b) for rod-like Hydra tissue fragments with non-zero spontaneous curvature. The chain rod shows resistance to both tangential extension and bending deformation. 
  } \label{Fig:Schematic-BeadSpring}
\end{figure}

We here introduce how to use the approximate variational method that is based on Onsager's' variational principle~\cite{Doi2021,Xu2021} to study the spontaneous bending dynamics of rod-like Hydra tissue fragments in two-dimensional $x-y$ plane (see Fig.~\ref{Fig:Schematic-BeadSpring}).

We first briefly explain Onsager's variational principle (OVP). Generally for a nonequilibrium system characterized by a set of state variables $\bm{\alpha} = (\alpha_1, \alpha_2,..., \alpha_{\rm f})$. The dynamics of the system described by the evolution of $\bm{\alpha}(t)$ with time $t$ is determined by minimizing a scalar function called Rayleighian that is defined as
\begin{equation}\label{eq:OVP-Ray} 
{\cal R}(\dot{\bm{\alpha}};\bm{\alpha}) = \Phi (\dot{\bm{\alpha}},\dot{\bm{\alpha}})+\dot{\cal F} (\dot{\bm{\alpha}};\bm{\alpha})
\end{equation}
with respect to the rates $\dot{\bm{\alpha}}$. Here $\Phi (\dot{\bm{\alpha}},\dot{\bm{\alpha}})\equiv \frac{1}{2}\xi_{ij}\dot{\alpha}_i\dot{\alpha}_j$ is the positive-definite dissipation-function with symmetric friction matrix satisfying $\xi_{ij}=\xi_{ji}$ and $\dot{\cal F}(\dot{\bm{\alpha}};\bm{\alpha}) \equiv\frac{\partial {\cal F}}{\partial \alpha_i}\dot{\alpha}_i$ is the change rate of free energy ${\cal F} ({\bm{\alpha}})$. Furthermore, if the state variables $\bm{\alpha}(t)$ can be written as some trial function of a small number of parameters denoted by $\bm{a}=\left(a_{1}, \, a_{2}, \ldots \right)$, \emph{i.e.}, $\bm{\alpha}(t)=\bm{\alpha}(\bm{a}(t))$, then the rate of the state variables $\dot{\bm{\alpha}}$ can be written as 
$\dot{\alpha}_{i}=\sum_{\rm m} \frac{\partial \alpha_{i}}{\partial a_{\rm m}} \dot{a}_{\rm m}$
and the Rayleighian becomes a function of the rates of parameters $\mathcal{R}(\dot{\bm{a}})$. The system dynamics approximated by the temporal evolution of the parameters $\bm{a}(t)$ can be determined by minimizing $\mathcal{R}$ with respect to $\dot{\bm{a}}$. 

The above Onsager's variational method provides a powerful tool of finding approximate solutions to the system dynamics and here we use it to study the spontaneous bending dynamics of thin Hydra rods that are modeled as continuous, compressible elastica curves~\cite{Haim2016} with contour length $\ell_{\rm c}$ and non-zero spontaneous curvature $c_0$. The spontaneous curvature radius is defined as $R_0=c_0^{-1}$ accordingly. In this case, the bending dynamics is descried by the evolution of the position vector $\bm{r}(s,t)=(x(s,t),y(s,t))$ of the elastica curve and the inter-layer sliding $\epsilon_{\rm s}(s,t)$ inside the composite Hydra tissue fragments. From the shape of the curve, we can calculate its total energy ${\cal F}_{\rm t}[\bm{r}(s,t),\epsilon_{\rm s}(s,t)]={\cal F}_{\rm c}+{\cal F}_{\rm s}+{\cal F}_{\rm b}$: the in-plane compression energy ${\cal F}_{\rm c}$ is included by
\begin{align}\label{eq:BendDyn-methods-Onsager-Fc}
{\cal F}_{\rm c}=\int_{-\ell_{\rm c}/2}^{\ell_{\rm c}/2} \frac{Y_{\rm r}}{2}\left(\lambda-1\right)^2 ds,
\end{align} 
the inter-layer sliding energy ${\cal F}_{\rm s}$, and the bending energy ${\cal F}_{\rm b}$ are, respectively, taken to be the phenomenological form of Eq.~(\ref{eq:Sliding-Rods-Ft0}) as 
\begin{align}\label{eq:BendDyn-methods-Onsager-Fsb} 
{\cal F}_{\rm s}=\int_{-\ell_{\rm c}/2}^{\ell_{\rm c}/2} \frac{Y_{\rm{s}}}{2}({\epsilon}_{\rm s}-{\epsilon}_{\rm s0}c/c_0)^{2} ds, \quad 
{\cal F}_{\rm b}=\int_{-\ell_{\rm c}/2}^{\ell_{\rm c}/2} \frac{D_{\rm r}}{2} \left(\frac{d \hat{\bm{t}}}{ds}-c_{0} \hat{\bm{n}}\right)^{2} ds=\int_{-\ell_{\rm c}/2}^{\ell_{\rm c}/2} \frac{D_{\rm r}}{2} \left(\frac{d\theta}{ds}-c_{0}\right)^{2}ds.
\end{align}
Here $\lambda=|{\partial \bm{r}(s,t)}/{\partial s}|$ is the local extension ratio, $\hat{\bm{t}}=\lambda^{-1}{\partial \bm{r}(s,t)}/{\partial s}=(\cos \theta(s), \sin \theta(s))$ is the local the unit tangential vector, $\hat{\bm{n}}$ is unit vector normal to $\hat{\bm{t}}$, and $\theta(s)\in[-\pi,\pi]$ is the local tangential orientation angle (as schematically shown in Fig.~\ref{Fig:Schematic-BeadSpring}). $Y_{\rm r}$ and $D_{\rm r}$ are the compression modulus and the flexural rigidity of the Hydra composite rod, respectively. 
Moreover, from the total energy ${\cal F}_{\rm t}[\bm{r}(s,t)]$ we obtain the rate of energy change $\dot{\cal F}_{\rm t}$ as a function of $\dot{\bm{r}}(s,t)\equiv d\bm{r}/dt$ and $\dot{\epsilon}_{\rm s}(s,t)\equiv d{\epsilon}_{\rm s}/dt$. 

In addition, to use OVP, we also need to find out the dissipation function of the rates $\dot{\bm{r}}(s,t)$. Since Hydra tissue fragments are bending in viscous liquid environment and the fragments are modeled as purely elastic materials, the dissipation during the bending arises from viscous drag in the surrounding fluid and the inter-layer frictional sliding inside the fragments. In this case, the dissipation function can be written as 
\begin{equation}\label{eq:BendDyn-methods-Onsager-Phi} 
\Phi=\int_{-\ell_{\rm c}/2}^{\ell_{\rm c}/2} ds \left(\frac{1}{2} \xi_{\rm v} \dot{\bm{r}}^{2}+\frac{1}{2} \xi_{\rm s} \dot{\epsilon}_{\rm s}^{2} \right),
\end{equation}  
with $\xi_{\rm v}$ and $\xi_{\rm s}$ being the viscous drag coefficient per rod length and friction coefficient for inter-layer sliding, respectively. 
Then minimizing Rayleighian ${\cal R}(\dot{\bm{r}};\bm{r}) = \Phi (\dot{\bm{r}},\dot{\bm{r}})+\dot{\cal F} (\dot{\bm{r}};\bm{r})$ with respect to $\dot{\bm{r}}$ gives a highly nonlinear partial differential equation (such as, Kirchhoff equations)~\cite{Callan2012}. 

In this work, we don't study this highly nonlinear equation that has been solved numerically extensively (particularly for very long thin rods)~\cite{Callan2012}, but we try, instead, to carry out approximate variational analysis introduced above by assuming some trial shape dynamics of the elastic curve. To make the assumed shape be smooth enough, we assume during the bending of the Hydra rod, its local curvature $c(s,t)$ takes some trial form (see Fig.~3 in the main text) of $c(s;\bm{a}(t))$ with $-\ell_{\rm c}/2 \leq s \leq \ell_{\rm c}/2$ being the arc-length parameter and $\bm{a}(t)$ denoting a set of parameters. From the trial local curvature $c(s;\bm{a}(t))$, we can calculate the local tangential orientation angle $\theta(s)$ by
\begin{subequations}\label{eq:BendDyn-methods-Onsager-thetatr}
\begin{equation} \label{eq:BendDyn-methods-Onsager-theta}
\theta(s,t)=\int_{0}^{s} c(\hat{s},t) d\hat{s}+\theta(0,t),
\end{equation}
from which we further calculate $\hat{\bm{t}}(s,t)$ and $\bm{r}(s,t)$, respectively, by  
\begin{equation}\label{eq:BendDyn-methods-Onsager-t}
\hat{\bm{t}}(s,t)=\frac{\partial \bm{r}(s,t)}{\partial s}/\left|\frac{\partial \bm{r}(s,t)}{\partial s}\right|=(\cos \theta(s,t), \sin \theta(s,t)),
\end{equation}
\begin{equation}\label{eq:BendDyn-methods-Onsager-r}
\bm{r}(s,t)=\left(\int_{0}^{s} \cos \theta(\hat{s},t) d\hat{s}+x(0,t), \int_{0}^{s} \sin \theta(\hat{s},t) d\hat{s}+y(0,t)\right).
\end{equation}
\end{subequations} 
Then using $\bm{r}(s,t)$ obtained from the trial curvature dynamics, we can calculate the Rayleighian $\cal R$ as a function of $\dot{\bm{a}}(t)$. Minimizing $\cal R$ with respect to $\dot{\bm{a}}(t)$ gives the ordinary differential equation for ${\bm{a}}(t)$, from which we solve the bending dynamics approximately. This type of calculations will be presented and discussed in Sec.~\ref{sec:BendDyn-results}. 

\subsubsection{Numerical simulations based on the bead-spring model for rods upon viscous dissipation \label{sec:BeadSpring}}

Now we introduce the bead-spring model (see Fig.~\ref{Fig:Schematic-BeadSpring}(b)) for the thin Hydra rod as an elastica curve with non-zero spontaneous curvature $c_0$. The bead-spring chain consists of $N$ point beads connected by $N-1$ linear springs of equilibrium length $\ell_0$. The total equilibrium contour length is then given by $\ell_{\rm c}=(N-1)\ell_0$. We assume that the chain rod shows resistance to both tangential (spring) deformation and bending deformation. Therefore, each bead in the chain is subjected to forces due to both spring deformation and bending (note that we here neglect the thermal fluctuating forces that are included in usual bead-spring models of polymers). Actually, this bead-spring model gives a discrete description of the continuous elastica curve considered in the above subsection.  

In the discrete bead-spring model, the two parts of the total energy, ${\cal F}_{\rm t}(\bm{r}_i(t))={\cal F}_{\rm c}+{\cal F}_{\rm b}$, are, respectively, given by 
\begin{equation}\label{eq:BendDyn-methods-BeadSpring-Fsb}
{\cal F}_{\rm c}=\sum_{i=1}^{N-1}\frac{Y_{\rm r}}{2\ell_0}\left(r_{i,i+1}-\ell_{0}\right)^{2}, \quad
{\cal F}_{\rm b}=\sum_{i=1}^{N-2} \frac{D_{\rm r}}{2 \ell_{0}}\left[\left(\hat{\bm{t}}_{i+1}-\hat{\bm{t}}_{i}\right)-\phi_{0 } \hat{\bm{n}}_{i}\right]^{2}. 
\end{equation}
with $\phi_0 \equiv \ell_0c_0=\ell_0/R_0$ being the equilibrium bending angle. 
Here ${r}_{i,i+1}=|\bm{r}_{i,i+1}|$ is the distance between neighbouring beads with $\bm{r}_{i,i+1} =\bm{r}_{i+1}-\bm{r}_i$ $(1\leq i\leq N-1)$ and $\bm{r}_i=(x_i,y_i)$ $(1\leq i\leq N)$ being the position vector of the $i$-th bead. Then the local extension ratio is given by  $\lambda={r}_{i,i+1}/\ell_0$. The local unit tangential vector and normal vector are given, respectively, by
\begin{equation}\label{eq:BendDyn-methods-BeadSpring-tn}
\hat{\bm{t}}_i  \equiv \frac{\bm{r}_{i,i+1}}{{r}_{i,i+1}}=(\cos \theta_i, \sin \theta_i), \quad
\hat{\bm{n}}_i=(-\sin \theta_i, \cos\theta_i), 
\end{equation}
in which $\theta_i$ (with $-\pi\leq \theta_i \leq \pi$) is the local tangential angle of each spring (defined relative to the horizontal $x$-direction, see Fig.~\ref{Fig:Schematic-BeadSpring}(b)).  
The angle $\phi_{i} \equiv \theta_{i+1}-\theta_{i}$ is the bending angle of $\hat{\bm{t}}_{i+1}$ relative to $\hat{\bm{t}}_i$. Note that $\phi_{i}$ can be either positive or negative; positive (negative) means bending upward (downward) as shown in Fig.~\ref{Fig:Schematic-BeadSpring}. Moreover, to be more accurate in the simulations for a rod with given contour length $\ell_{\rm c}=(N-1)\ell_0$, we need to choose a large $N$ and hence a small $\ell_0$ such that $|\phi_i| \ll 1$. When $\theta_{i+1} \leq -\pi/2$ (third quadrant) and $\theta_{i} \geq  \pi/2$ (second quadrant) may lead to a large value of $|\phi_i|$. To ensure the validity of the bead-spring model, we define $\phi_i=\theta_{i+1}-\theta_{i}+2\pi$  in this case. 

From the total energy ${\cal F}_{\rm t}(\bm{r}_i(t))$ we can obtain the rate of energy change $\dot{\cal F}_{\rm t}$ as a function of $\dot{\bm{r}}_i(t)\equiv d\bm{r}_i/dt$. The dissipation function in Eq.~(\ref{eq:BendDyn-methods-Onsager-Phi}) reduces to 
\begin{equation}\label{eq:BendDyn-methods-BeadSpring-Phi}
\Phi(\dot{\bm{r}}_i)=\sum_{i=1}^{N} \frac{1}{2} \xi_{\rm b}  \dot{\bm{r}}_i(t)^{2},
\end{equation}
with $\xi_{\rm b} \equiv \xi_{\rm v}\ell_0$ being the effective drag coefficient of the chain segments in the surrounding viscous fluids. Then minimizing Rayleighian ${\cal R}(\dot{\bm{r}}_i) = \Phi (\dot{\bm{r}}_i)+\dot{\cal F}_{\rm t} (\dot{\bm{r}}_i)$ with respect to $\dot{\bm{r}}_i$ gives the dynamic equations of the bead-spring chain as
\begin{equation}\label{eq:BendDyn-methods-BeadSpring-DynEqn}
\xi_{\rm b} \frac{d\bm{ r}_{i}}{d t}=\bm{F}_{i}^{\rm{s}}+\bm{F}_{i}^{\rm{b}}, \quad i=1,2, \ldots N, 
\end{equation}
in which the spring force $\bm{F}_{i}^{\rm{s}}\equiv -{\partial {\cal F}_{\rm c}}/{\partial\bm{r}_{i}}$ is given by
\begin{subequations}\label{eq:BendDyn-methods-BeadSpring-Fs}
\begin{align}
\bm{F}_{i}^{\rm{s}}=&-\frac{Y_{\rm r}}{\ell_0}\left[\left(r_{i-1, i}-\ell_{0}\right)\hat{\bm{t}}_{i-1}-\left(r_{i, i+1}-\ell_{0}\right) \hat{\bm{t}}_i\right], \quad 2\leq i \leq N-1, \\
& \bm{F}_{1}^{\rm{s}}=\frac{Y_{\rm r}}{\ell_0}\left(r_{1,2}-\ell_{0}\right)\hat{\bm{t}}_{1}, \quad 
\bm{F}_{N}^{\rm{s}}=-\frac{Y_{\rm r}}{\ell_0}\left(r_{N-1,N}-\ell_{0}\right)\hat{\bm{t}}_{N-1}, 
\end{align}  
and the bending force $\bm{F}_{i}^{\rm{b}}\equiv -{\partial {\cal F}_{\rm b}}/{\partial \bm{r}_{i}}$ is given by  
\begin{align} \label{eq:BendDyn-methods-BeadSpring-Fb}
&\quad \quad \bm{F}_{i}^{\rm {b}}=\frac{D_{\rm r}}{\ell_0}\left[\left(\Delta \phi_{i-1}-\Delta \phi_{i-2}\right) \frac{\hat{\bm{n}}_{i-1}}{r_{i-1, i}}-\left(\Delta \phi_{i}-\Delta \phi_{i-1}\right) \frac{\hat{\bm{n}}_{i}}{r_{i, i+1}}\right], \quad 3 \leq i \leq N-2, \\
&\quad \quad \quad \bm{F}_{1}^{\rm {b}}=-\frac{D_{\rm r}}{\ell_0} \Delta \phi_1 \frac{\hat{\bm{n}}_{1}}{r_{1,2}}, \quad \quad \bm{F}_{2}^{\rm {b}}=\frac{D_{\rm r}}{\ell_0} \left[\Delta \phi_{1}\frac{\widehat{\bm{n}}_{1}}{r_{1,2}}-\left(\Delta \phi_{2}-\Delta \phi_{1}\right) \frac{\widehat{\bm{n}}_{2}}{r_{2,3}}\right], \\
&\bm{F}_{N-1}^{\rm {b}}=\frac{D_{\rm r}}{\ell_0}\left[\left(\Delta \phi_{N-2}-\Delta \phi_{N-3}\right) \frac{\widehat{\bm{n}}_{N-2}}{r_{N-2, N-1}}+\Delta \phi_{N-2} \frac{\widehat{\bm{n}}_{N-1}}{r_{N-1, N}}\right], \quad 
\bm{F}_{N}^{\rm {b}}=-\frac{D_{\rm r}}{\ell_0} \Delta \phi_{N-2} \frac{\widehat{\bm{n}}_{N-1}}{r_{N-1, N}},
\end{align} 
\end{subequations}
with $\Delta \phi_i\equiv (\sin{\phi_{i}}-\phi_{0}\cos\phi_{i})$ characterizing the deviation of bending angle $\phi$ from equilibrium $\phi_0$. 
Here the bending force is obtained by using 
\begin{subequations}\label{eq:BendDyn-methods-BeadSpring-Forceb123}
\begin{equation}\label{eq:BendDyn-methods-BeadSpring-Forceb1}
\bm{F}_{i}^{\rm{b}}=-\frac{D_{\rm r}}{\ell_{0}}\Delta \phi_j\frac{\partial \phi_{j}}{\partial \bm{r}_{i}}=\frac{D_{\rm r}}{\ell_{0}} \frac{\Delta \phi_j}{\sin \phi_{j}} \frac{\partial\left(\hat{\bm{t}}_{j+1} \cdot \hat{\bm{t}}_{j}\right)}{\partial \bm{r}_{i}}, 
\end{equation} 
and the following identities
\begin{equation}\label{eq:BendDyn-methods-BeadSpring-Forceb2}
\frac{d r_{j,j+1}}{d \bm{r}_{i}}=\hat{\bm t}_{j}  \left(\delta_{j+1, i}-\delta_{j i}\right), 
\quad 
\frac{\partial \hat{\bm t}_{k} }{\partial \bm{r}_{i}}=\frac{\bm{I} -\hat{\bm t}_{k} \hat{\bm t}_{k}}{r_{k,k+1}} \left(\delta_{k+1, i}-\delta_{k i}\right), 
\end{equation}
\begin{equation}\label{eq:BendDyn-methods-BeadSpring-Forceb3}
\hat{\bm t}_{k+1}\cdot (\bm{I} -\hat{\bm t}_{k} \hat{\bm t}_{k})=\sin \phi_k\hat{\bm n}_{k}, 
\quad 
\hat{\bm t}_{k}\cdot (\bm{I} -\hat{\bm t}_{k+1} \hat{\bm t}_{k+1})=-\sin \phi_k\hat{\bm n}_{k+1},  
\end{equation}
\end{subequations}
which can be obtained easily from Eq.~(\ref{eq:BendDyn-methods-BeadSpring-tn}). Furthermore, note that $\phi_i, \l \phi_0 \ll 1$ and hence $\Delta \phi_i \approx \phi_{i}-\phi_{0}$. Then the bending energy in Eq.~(\ref{eq:BendDyn-methods-BeadSpring-Fsb}) reduces to
\begin{equation}\label{eq:BendDyn-methods-BeadSpring-Fb2}
{\cal F}_{\rm{b}}(\theta_i) \approx \sum_{i=1}^{N-2} \frac{D_{\rm r}}{2\ell_{0}}\left(\phi_{i}-\phi_{0}\right)^{2},
\end{equation}
from which we find the bending force 
\begin{equation}\label{eq:BendDyn-methods-BeadSpring-Forceb20} 
\bm{F}_{i}^{\rm{b}}= 
-\frac{D_{\rm r}}{\ell_{0}}\left(\phi_{j}-\phi_{0}\right) \frac{\partial \phi_{j}}{\partial \bm{r}_{i}}=\frac{D_{\rm r}}{\ell_{0}} \frac{\phi_{j}-\phi_{0}}{\sin \phi_{j}} \frac{\partial\left(\hat{\bm{t}}_{j+1} \cdot \hat{\bm{t}}_{j}\right)}{\partial \bm{r}_{i}},  
\end{equation}  
which can also be obtained directly from Eq.~(\ref{eq:BendDyn-methods-BeadSpring-Forceb1}). 

In addition, the dynamic equation (\ref{eq:BendDyn-methods-BeadSpring-DynEqn}) for the bead-spring chain can be dedimensionalized by taking the unit of length by $R_0$, the time by $t_{0}=\xi_{\rm b} R_{0}^{3}/D_{\rm r}$, and the force by $F_{0}=D_{\rm r}/R_0^2$. In dimensionless forms, the dynamic equation of the bead-spring model is given by
\begin{equation}\label{eq:BendDyn-methods-BeadSpring-DynEqn2}
\frac{d \tilde{\bm{r}}_{i}}{d \tilde{t}}=\widetilde{\bm{F}}_{i}^{\rm{s }}+\widetilde{\bm{F}}_{i}^{\rm{b}}, 
\end{equation}
in which the dimensionless forces are given, for example, by
\begin{subequations}\label{eq:BendDyn-methods-BeadSpring-Forcesb2}
\begin{equation}\label{eq:BendDyn-methods-BeadSpring-Forces2}
\widetilde{\bm{F}}_{i}^{\rm{s}}=-{\mathcal{K}}_{\rm s}\left[\left(\tilde{r}_{i-1, i}-\phi_0\right) \frac{\tilde{\bm{r}}_{i-1, i}}{\tilde{r}_{i-1, i}}-\left(\tilde{r}_{i, i+1}-\phi_0\right) \frac{\tilde{\bm{r}}_{i, i+1}}{\tilde{r}_{i, i+1}}\right], 
\end{equation}
\begin{equation}\label{eq:BendDyn-methods-BeadSpring-Forceb2}
\widetilde{\bm{F}}_{i}^{\rm{b}}= \frac{1}{\phi_0}\left[(\Delta \phi_{i-1}-\Delta \phi_{i-2}) \frac{\widehat{\bm{n}}_{i-1}}{\tilde{r}_{i-1, i}}-\left(\Delta \phi_{i}-\Delta \phi_{i-1}\right) \frac{\widehat{\bm{n}}_{i}}{\tilde{r}_{i, i+1}}\right]. 
\end{equation}
\end{subequations}
Three dimensionless parameters appear in the above equations: (1) Effective spring stiffness $\mathcal{K}_{\rm s}={Y_{\rm r} R_{0}^2}/{D_{\rm r}\ell_0}$; (2) Equilibrium bending angle $\phi_0 \equiv \ell_0c_0=\ell_0/R_0$; (3) Number of beads $N=\ell_{\rm c}/\ell_0+1$, and hence the normalized contour length is given by $\tilde{\ell}_{\rm c}\equiv c_0\ell_c=\ell_{\rm c}/R_0=(N-1)\phi_0$. In our simulations, we take (if no additional explanations) $\phi_0=\pi/30$ and the normalized contour length $\tilde{\ell}_{\rm c}$ varying from $\pi/4$ (the perimeter of $1/8$ circle) to $4\pi$ (the perimeter of two circles). Particularly, we set $\mathcal{K}_{\rm s}=1000$ (\emph{i.e.}, a very large $Y_{\rm r}$) so that the distances ${r}_{i,i+1}$ between neighbouring beads almost do not change with time and equal to the equilibrium constant $\ell_0$. 

\begin{figure}[htbp]
  \centering
  \includegraphics[width=0.8\textwidth]{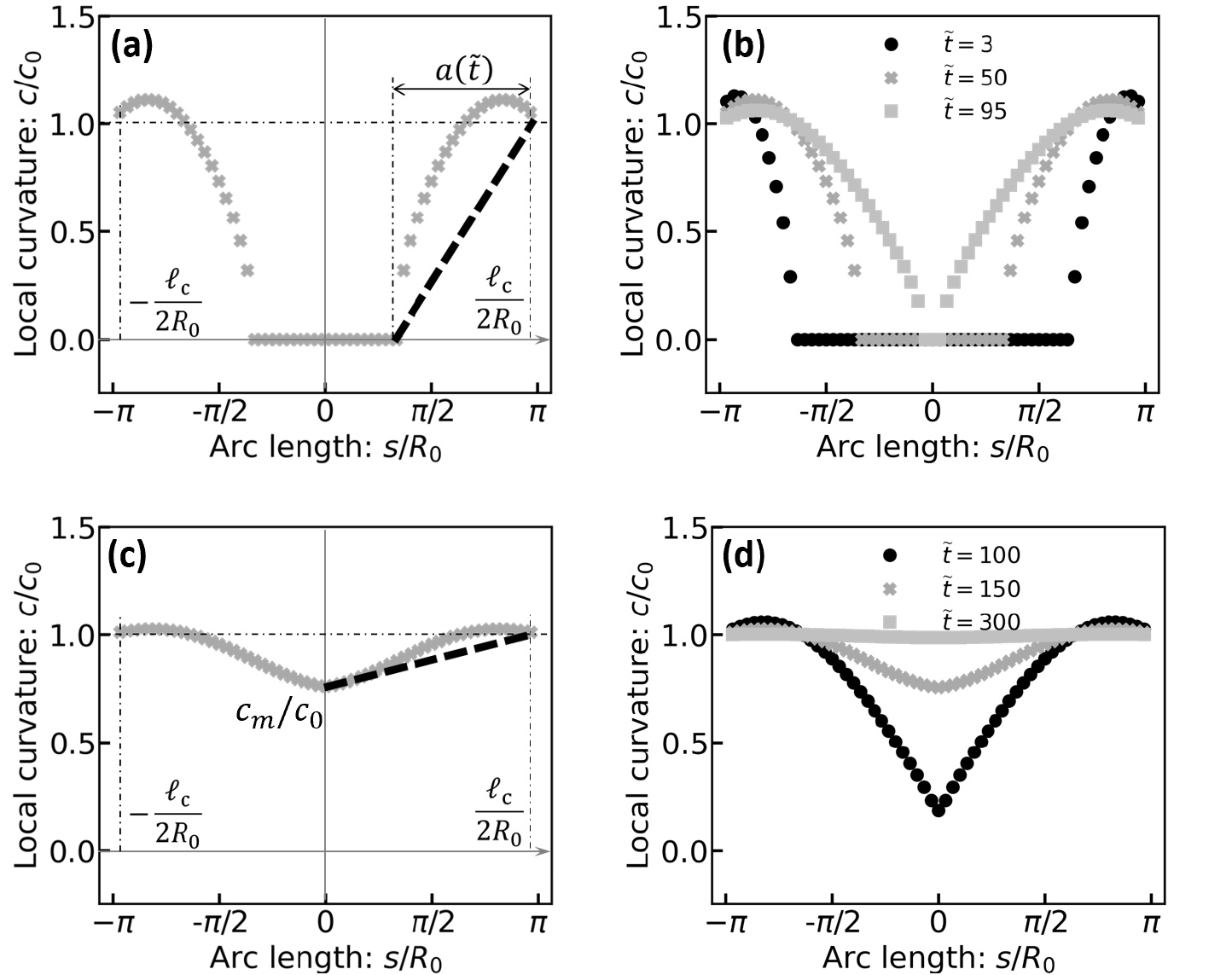}
  \caption {Shape of a Hydra rod-like fragment (of contour length $\ell_{\rm c}=2\pi R_0$) obtained from the bead-spring model during its spontaneous bending in viscous fluids on a (non-sticky) supporting surface (Supp. B.C.). 
  The Hydra rod bends spontaneously from initial flattened state as cut from Hydra body. The bending process can be simply divided into two stages: (i) the initial bending stage that starts from the edges and propagates diffusively into the center as shown in (a) and (b) at several different time $\tilde{t}=t/t_0$; (ii) the final bending stage that is close to the equilibrium bent state with homogeneous curvature $c_0=R_0^{-1}$ as shown in (c) and (d) at several $\tilde{t}$. In both stages, a one-parameter linear curvature profile (shown as dashed lines and parameterized by $a(\tilde{t})$ and $c_{\rm m}(\tilde{t})$, respectively) is assumed in order to carry out variational analysis as done in Sec.~\ref{sec:BendDyn-results}.
  } \label{Fig:Schematic-BeadSpring1}
\end{figure}

\subsection{Two regimes of the bending dynamics }\label{sec:BendDyn-results}

\begin{figure}[htbp]
  \centering
  \includegraphics[width=0.8\textwidth]{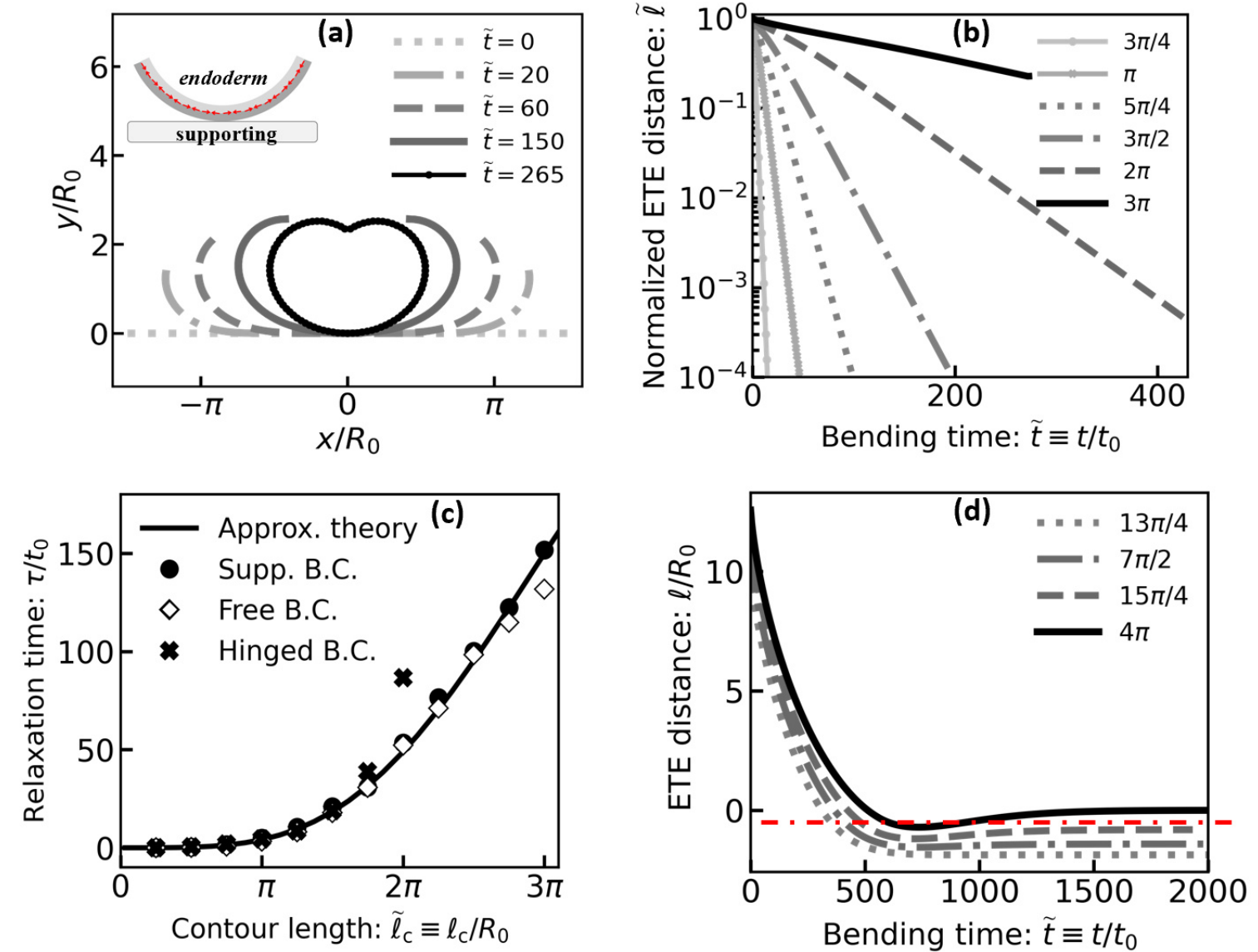}
  \caption {Dynamic bending process of a Hydra rod-like fragment in viscous fluids on a (non-sticky) supporting surface. (a) The shape evolution of the Hydra rod (of contour length $\ell_{\rm c}=3\pi R_0$) obtained numerically from the bead-spring model. (b,d) The temporal evolution of the normalized end-to-end (ETE) distance $\tilde{\ell}$ (defined in Eq.~(\ref{eq:BendDyn-results-regime1-ETE})) follows power-law relations in the initial bending stage (as shown in Fig.~3(a) by a log-log plot in the main text), and an exponential relation in the final bending stage close to the equilibrium bent state (as shown in (b) by a semi-log plot). 
(c) The characteristic time $\tau$ for the exponential decay of $\tilde{\ell}$ shows a universal and nonlinear dependence on the contour length $\tilde{\ell}_{\rm c}$ of the tissue rod for three different boundary conditions.
Here the rod contour length ($\tilde{\ell}_{\rm c}\equiv \ell_{\rm c}/R_0$) varies from $3\pi/4$ to $3\pi$. (d) For longer Hydra rods with $\tilde{\ell}_{\rm c}$ varying from $13\pi/4$ to $4\pi$, the ETE (by linear plot) becomes negative and is non-monotonic in the final stage after the two edges of the rod cross each other.
  } \label{Fig:Schematic-BeadSpring2}
\end{figure}

We now present the major theoretical results that we obtain for the bending dynamics of rod-like Hydra tissue fragments (with spontaneous curvature $c_0$) in viscous fluid environment using the methods introduced above: the approximate variational analysis of a continuous elastica curve and the numerical computations of the discrete bead-spring model. We focus on the bending dynamics of short rod-like fragments (with contour length $\ell_{\rm c} \leq 2\pi R_0$ or $\tilde{\ell}_{\rm c}\leq 2\pi$) starting from the flattened state on a (non-sticky) supporting surface (see Fig.~\ref{Fig:Schematic-BeadSpring2}(a) for the simulations of the bending process). We found that the bending process starts from the edges and propagates into the middle region. Two major regimes (initial diffusive bending regime and final relaxational bending regime) are identified according to how far it is from the final equilibrium bent state with uniform curvature $c_0$ and equilibrium end-to-end (ETE) distance $\ell_{\rm eq}$. 

Before presenting our calculations, we would like to give two general remarks. (i) In general, the curvature of a single elastica curve is always continuous everywhere unless there are some point torques/forces applied on the curve. (ii) The bending dynamics of the rod we considered here has a mirror symmetry with respect to $s=0$ (for example, see Fig.~\ref{Fig:Schematic-BeadSpring1} and Fig.~\ref{Fig:Schematic-BeadSpring2}(a)) so that hereafter we do the calculations for \emph{right-half} of the rod, $s\in [0, \ell_{\rm c}/2]$. Moreover, from symmetry considerations, we have $\theta(0,t)=0$, and particularly for Hydra tissue fragments on supporting boundary, we have $\bm{r}(0,t)=0$. 



\subsubsection{Relaxational bending dynamics close to the equilibrium bent state}\label{sec:BendDyn-results-regime1}

In the final stage of the bending dynamics close to the equilibrium bent state, the local curvature shown in Fig.~\ref{Fig:Schematic-BeadSpring1}(c,d) as a function of arc-length is obtained from numerical simulations using the bead-spring model. It can be seen that the curvature changes continuously from some non-zero curvature $c_{\rm m}(t)$ (that changes with time) at the middle position $s=0$ to spontaneous curvature $c_0$ at the edge, $s=\pm \ell_{\rm c}/2$. As introduced in Sec.~\ref{sec:BendDyn-methods-Onsager}, we should assume a trial shape dynamics that can be described by a trial time-varying local curvature $c(s, t)$ to carry out direct variational analysis. It is a good approximation to assume that $c(s, t)$ takes the following linear profile:
\begin{equation}\label{eq:BendDyn-results-regime1-c}
c(s, t)=c_0\left(1-r_{\rm c}+\frac{2 s}{\ell_{\rm c}} r_{\rm c} \right), \quad 0 \leq  s \leq \ell_{\rm c},
\end{equation}
with $r_{\rm c}(t)\equiv 1-c_{\rm m}(t)/c_0$.  
For the potential slide between layers, we also assume a linear profile as
\begin{equation} \label{eq:Sliding-Rods-es} 
\epsilon_{\rm s}=\epsilon_{\rm s 0}\left(1-r_{\rm s}+\frac{2 s}{\ell_{\rm c}} r_{\rm s} \right), \quad 0 \leq  s \leq \ell_{\rm c},
\end{equation} 
with $r_{\rm s}=1-\epsilon_{\rm s,m} /\epsilon_{\rm s 0}$ and $\epsilon_{\rm s,m}$ being the slide at the symmetry position $s=0$. We think this assumption is reasonable because it is the curvature that induces the in-plane sliding between layers. Given the profile of $c(s,t)$ and $\epsilon_{\rm s}(s,t)$ assumed above, the bending dynamic process is then represented by two parameters $r_{\rm c}(t)$ and $r_{\rm s}(t)$. Note that near the final equilibrium bent state, we have $c_{\rm m}(t)\to c_0$, $\epsilon_{\rm s,m} (t)\to \epsilon_{\rm s 0}$, and hence $r_{\rm c}(t), \, r_{\rm s}(t) \ll 1$.  

Substituting the trial curvature profile in Eq.~(\ref{eq:BendDyn-results-regime1-c}) into Eqs.~(\ref{eq:BendDyn-methods-Onsager-thetatr}) and using the symmetric boundary condition $\phi(0)=0$ and supporting boundary condition $\bm{r}(0,t)=0$, we obtain the bending angle, tangent unit vector, and position vector of beads as
\begin{subequations}\label{eq:BendDyn-results-regime1-phitaur}
\begin{equation}\label{eq:BendDyn-results-regime1-phi}
\phi(s)=c_0(1-r_{\rm c})s+\frac{r_{\rm c}c_0}{\ell_{\rm c}}s^2, 
\end{equation}
\begin{equation}\label{eq:BendDyn-results-regime1-tau}
\hat{\bm{t}}(s,t)\approx  
\left(\cos (c_0s)+r_{\rm c}c_{0} \left(s-\frac{s^2}{\ell_{\rm c}}\right)  \sin( c_{0} s), \quad \sin (c_{0} s)-r_{\rm c}c_{0} \left(s-\frac{s^2}{\ell_{\rm c}}\right)  \cos (c_{0} s) \right),
\end{equation}
\begin{equation}\label{eq:BendDyn-results-regime1-r}
\begin{aligned}
&c_0 x(s,t)\approx \sin (c_{0} s)+\frac{1}{\tilde{\ell}_{\rm c}}\left[2+\left(-2-c_{0}\tilde{\ell}_{\rm c} s+c_{0}^{2} s^{2}\right) \cos (c_{0} s)+\left(\tilde{\ell}_{\rm c}-2 c_{0} s\right) \sin (c_{0} s)\right] r_{\rm c}, \\
&c_0 y(s,t)\approx 1-\cos (c_{0} s) +\frac{1}{\tilde{\ell}_{\rm c}}\left[\tilde{\ell}_{\rm c}+\left(-\tilde{\ell}_{\rm c}+2 c_{0} s\right) \cos (c_{0} s)-\left(2+c_{0}\tilde{\ell}_{\rm c} s-c_{0}^{2} s^{2}\right) \sin (c_{0} s)\right] r_{\rm c}, 
\end{aligned}
\end{equation}
\end{subequations}
respectively. Using Eq.~(\ref{eq:BendDyn-results-regime1-r}), we obtain the ETE distance by $\ell(t)=2x\left(s={\ell_{\rm c}}/{2}, t\right)$, from which we calculate the normalized ETE distance by 
\begin{equation}\label{eq:BendDyn-results-regime1-ETE}
\tilde{\ell}\equiv \frac{\ell-\ell_{\rm eq}}{\ell_{\rm c}-\ell_{\rm eq}} = 
\frac{1-\left(1+ \tilde{\ell}_{\rm c}^{2}/8 \right) \cos (\tilde{\ell}_{\rm c}/2)}{\tilde{\ell}_{\rm c}^2/4-(\tilde{\ell}_{\rm c}/2)\sin (\tilde{\ell}_{\rm c}/2)} r_{\rm c}
\equiv \mathcal{A}^{-1} r_{\rm c}. 
\end{equation}
Here $c_0\ell_{\rm eq}=2\sin (\tilde{\ell}_{\rm c}/2)$ is the normalized equilibrium ETE distance, $\tilde{\ell}_{\rm c}\equiv c_0\ell_c=\ell_{\rm c}/R_0$ is the normalized contour length, and $\mathcal{A}$ is defined to simplify the notation. Note that initially at the flattened state, we have $\ell=\ell_{\rm c}$ and hence $\tilde{\ell}=1$. At the final equilibrium state, we have $\ell=\ell_{\rm eq}$ and hence $\tilde{\ell}=0$. 

The bending energy and dissipation function of the rod fragment can then be obtained from  Eqs.~(\ref{eq:BendDyn-methods-Onsager-Fsb}) and (\ref{eq:BendDyn-results-regime1-phi}) as 
\begin{subequations}\label{eq:BendDyn-results-regime1-FtPhi}
\begin{equation}\label{eq:BendDyn-results-regime1-Ft}
\mathcal{F}_{\rm t}=\frac{1}{12} Y_{\rm{s}} \ell_{c} \epsilon_{\rm s 0}^2\left(r_{\rm c}-r_{\rm s}\right)^{2}+\frac{1}{12}D_{\rm{r}} \ell_{c}  c_{0}^{2} r_{\rm c}^{2},
\end{equation}
\begin{equation}\label{eq:BendDyn-results-regime1-Phi}
\Phi=\frac{1}{2} \xi_{\rm v} \ell_{\rm c} c_{0}^{-2} \mathcal{L} \dot{r}_{c}^{2}+\frac{1}{12} \xi_{\rm s} \ell_{\rm c}  \epsilon_{\rm s 0}^{2} \dot{r}_{\rm s}^{2},
\end{equation}
\end{subequations}
with $\mathcal{L}(\tilde{\ell}_{\rm c})\equiv 1+\frac{1}{60}\tilde{\ell}_{\rm c}^{2}+\frac{1}{2}\left(1 +\frac{24}{\tilde{\ell}_{\rm c}^2}\right)\left[\cos (\tilde{\ell}_{\rm c}/2)-\frac{2}{\tilde{\ell}_{\rm c}}\sin (\tilde{\ell}_{\rm c}/2)\right]$. Note that when the contour length of the rod tends to zero, $\tilde{\ell}_{\rm c} \to 0$, we have $\mathcal{L}\to 0$, that is, $\Phi\to 0$. Minimizing the Rayleighian $\mathcal{R}=\dot{\cal F}_{\rm{t}}+\Phi$ with respect to the rates $\dot{r}_{\rm c}$ and $\dot{r}_{\rm s}$ yields the kinetic equations:
\begin{align} \label{eq:Sliding-results-regime1-DynEqn}
\tau \dot{r}_{\rm c}=-(1+{\cal B}) r_{\rm c}+{\cal B} r_{\rm s}, \quad \tau_{\rm s} \dot{r}_{\rm s}=r_{\rm c}-r_{\rm s}, 
\end{align}
where ${\cal B}={Y_{\rm{s}} \epsilon_{\rm s0}^{2}}/{D_{\rm{r}} c_{0}^{2}}$, $\tau \equiv {6 \xi_{\rm v}\mathcal{L}}/{D_{\rm{r}} c_0^4}$, and $\tau_{\rm s} \equiv {\xi_{\rm s}}/{Y_{\rm{s}}}$. 

The general solution of Eq.~(\ref{eq:Sliding-results-regime1-DynEqn}) can be easily obtained. However, we think the following two limits are more interested and relevant to the bending dynamics during tissue morphogenesis. 
\begin{itemize}
\item {\bf{Limits of coherent Hydra rods or incoherent Hydra rods with small friction and $\tau_{\rm s}/\tau \ll 1$}}.  In this case, the bending dynamics is described by only one parameter, ${r}_{\rm c}(t)$. Then from Eq.~(\ref{eq:Sliding-results-regime1-DynEqn}) we obtain the bending dynamics described by
\begin{align}\label{eq:Sliding-results-regime1-DynEqn}
\tau \dot{r}_{\rm c}=- r_{\rm c}, 
\end{align} 
and hence from Eq.~(\ref{eq:BendDyn-results-regime1-ETE}) we get
\begin{equation}\label{eq:BendDyn-results-regime1-ETEt}
\tilde{\ell} \sim \exp\left(-t/\tau\right).
\end{equation}
Thus, the dimensionless characteristic relaxational time for bending close to the equilibrium is given by
\begin{equation}
\tau/t_0 = \frac{6}{\phi_0} \mathcal{L}(\tilde{\ell}_{\rm c}), \quad 
{\rm or},\quad \tau=\frac{6 \xi_{\rm v}}{D_{\rm{r}} c_0^4} \mathcal{L}(\tilde{\ell}_{\rm c}).
\end{equation}
with $\phi_0=\pi/30$ in the simulations. 
Note that $\tau$ depends on $\mathcal{L}(\tilde{\ell}_{\rm c})$ that shows nonlinear dependence on the the contour length $\ell_{\rm c}$ of the rod: $\tau$ is very small for $\ell_{\rm c}<\pi R_0$, but increases very fast with increasing $\ell_{\rm c}$ when $\ell_{\rm c}>\pi R_0$ as shown in Fig.~\ref{Fig:Schematic-BeadSpring2}(c). For example, $\tau$ for $\ell_{\rm c}=2\pi R_0$ is $\sim 10$ times larger than $\tau$ for $\ell_{\rm c}=\pi R_0$. 

\item {\bf{Limit of incoherent Hydra rods with large friction and $\tau_{\rm s}/\tau \gg 1$}}. If the elastic Hydra rod is incoherent and the interlayer friction coefficient $\xi_{\rm s}$ is very large with $\tau_{\rm s}/\tau \gg 1$, then from Eq.~(\ref{eq:Sliding-results-regime1-DynEqn}) we find 
\begin{align} 
(1+{\cal B}) r_{\rm c}\approx {\cal B} r_{\rm s}, \quad \tau_{\rm s} (1+{\cal B}) \dot{r}_{\rm c}=-r_{\rm c}, 
\end{align}
and the solution is 
\begin{align}
r_{\rm c} \sim \exp\left[-t/\tau_{\rm s} (1+{\cal B})\right], \quad
r_{\rm s}\approx(1+{\cal B}^{-1}) r_{\rm c}. 
\end{align}
From Eq.~(\ref{eq:BendDyn-results-regime1-ETE}), we obtain the normalized ETE distance as
\begin{equation} 
\tilde{\ell}= \mathcal{A}^{-1} r_{\rm c}\sim \exp\left[-t/\tau_{\rm s} (1+{\cal B})\right]. 
\end{equation}

\end{itemize}

\subsubsection{Diffusive bending dynamics around the initial flattened state} 

Now we consider the initial stage of the bending dynamics starting from the flattened state on the supporting surface (see Fig.~\ref{Fig:Schematic-BeadSpring2} for the simulations of the bending process). In this stage, the local curvature $c(s,t)$ and the ETE distance $\ell \approx \ell_{\rm c}$ is computed from numerical simulations and shown in Fig.~\ref{Fig:Schematic-BeadSpring1} and Fig.~\ref{Fig:Schematic-BeadSpring2}, respectively. It is found that $c(s,t)$ changes continuously from zero in the middle flat region (with $c=0$) around $s=0$ to non-zero in the curved region near the edge, and to the spontaneous curvature $c_0$ at the edge, $s=\pm \ell_{\rm c}/2$. 

As done previously, we carry out direct variational analysis starting from a trial profile of time-varying local curvature $c(s, t)$. The simplest trial curvature profile that is continuous and change between $0$ at $s=0$ and $c_0$ at the edge $s=\ell_{\rm c}/2$ would be the following linear function
\begin{equation}\label{eq:BendDyn-results-regime2-c}
c(s;a(t))=
  \begin{cases}
  0,      &-\ell_{\rm c} / 2+a \leq \bar{s} < 0, \\
  {c_{0}}\bar{s}/a, & \quad \quad 0 \leq \bar{s} \leq a, 
  \end{cases}
\end{equation}  
with $\bar{s} \equiv s-\ell_{\rm c} / 2+a(t)$ being a convenient auxiliary variable. Here in the trial bending dynamics described by $c(s;a(t))$, $a(t)$ is the length of the bent part of the rod with non-zero curvature and it is the only parameter changing with time.

At the very beginning of the bending process, the normalized length $ac_0$ of the bent part is very small, that is $r_{\rm c}(t)c_0\ll 1$. Substituting the trial curvature profile in Eq.~(\ref{eq:BendDyn-results-regime2-c}) into Eqs.~(\ref{eq:BendDyn-methods-Onsager-thetatr}) and using the symmetric boundary condition $\phi(\bar{s}=0)=0$ and supporting boundary condition $\bm{r}(\bar{s}=0)=(\ell_{\rm c} / 2-a(t),0) $, we obtain the position vector of beads as
\begin{equation} 
\bm{r}(\bar{s})=\left(\frac{a}{\tilde{a}} {\cal C}\left({\tilde{a}}{\bar{s}}/a\right)+\frac{\ell_{\rm c}}{2}-a, \quad \frac{a}{\tilde{a}}  {\cal S} \left({\tilde{a}}{\bar{s}}/a \right)\right), \quad  0\leq  \bar{s} \leq a,
\end{equation}
in which $\tilde{a}(t) \equiv \sqrt{a(t)c_{0}/\pi}$, the two functions ${\cal C}(x)$ and ${\cal S}(x)$ are called Fresnel integrals. The ETE distance is then given by
\begin{equation}
\ell=2 x(\bar{s}=a)=2\frac{a}{\tilde{a}}{\cal C} \left({\tilde{a}}\right)+\ell_{\rm c}-2a, 
\end{equation} 
and using $r_{\rm c}c_0\ll 1$ (hence $\tilde{a}\ll 1$) we find the normalized ETE distance $\tilde{\ell}$ satisfies
\begin{equation}\label{eq:mtob} 
1-\tilde{\ell}= \frac{\ell_{\rm c}-\ell}{\ell_{\rm c}-\ell_{\rm eq}} \approx \frac{a^3}{20(\ell_{\rm c}-\ell_{\rm eq})}. 
\end{equation}
The bending energy  and the dissipation function of the rod fragment can then be obtained from Eqs.~(\ref{eq:BendDyn-methods-Onsager-Fsb}) and (\ref{eq:BendDyn-results-regime1-phi}) as 
\begin{equation}
{\cal F}_{\rm{b}}=\int_{-\frac{\ell_{\rm c}}{2}+a}^{0} \frac{D_{\rm r}}{2}c_{0}^{2} d \bar{s} + \int_{0}^{a} \frac{D_{\rm r}}{2}\left(\frac{c_{0}}{a}\bar{s}-c_{0}\right)^{2} d \bar{s}\approx \frac{1}{2}D_{\rm r} c_{0}^{2}\left(\frac{\ell_{\rm c}}{2} -\frac{2a}{3} \right), \quad
\Phi=\int_{0}^{a} \frac{1}{2}  \xi_{\rm v} \dot{\bm{r}}^{2} d \bar{s}=\frac{1}{2}\xi_{\rm v}{\cal Z} a\dot{a}^{2},
\end{equation} 
respectively, in which  
\begin{equation}
\begin{aligned}
{{\cal Z}}=\frac{4}{3}+\frac{1}{\tilde{a}^{2}} \left\{ 
 {\cal C}^2(\tilde{a})+{\cal S}^2(\tilde{a}) +\frac{{\cal S}(\tilde{a})}{{\tilde{a}}}\left[ \cos(\tilde{a}^{2})+\frac{2}{\pi}\cos\left(\frac{\pi\tilde{a}^{2}}{2}\right)\right] - \sqrt{\frac{\pi}{\pi-2}}\frac{{\cal S}(\tilde{a}\sqrt{1-\pi/2})}{{\tilde{a}}} \right.     \\
\left.  -\frac{{\cal C}(\tilde{a})}{{\tilde{a}}}\left[ \sin(\tilde{a}^{2})+\frac{2}{\pi}\sin\left(\frac{\pi\tilde{a}^{2}}{2}\right)\right]
+2{{\tilde{a}}}{{\cal C}(\tilde{a})}
-\sin(\tilde{a}^{2})-\frac{2}{\pi}\sin \left(\frac{\pi\tilde{a}^{2}}{2}\right) \right \},
\end{aligned}
\end{equation}
and we find ${{\cal Z}} \approx 1+0.12 \tilde{a}^{4}$ to the fourth order of $\tilde{a}\ll 1$ and hence $\Phi \approx \frac{1}{2} \xi_{\rm v} a\dot{a}^{2}$. 

\begin{figure}[htbp]
  \centering
  \includegraphics[width=0.8\columnwidth]{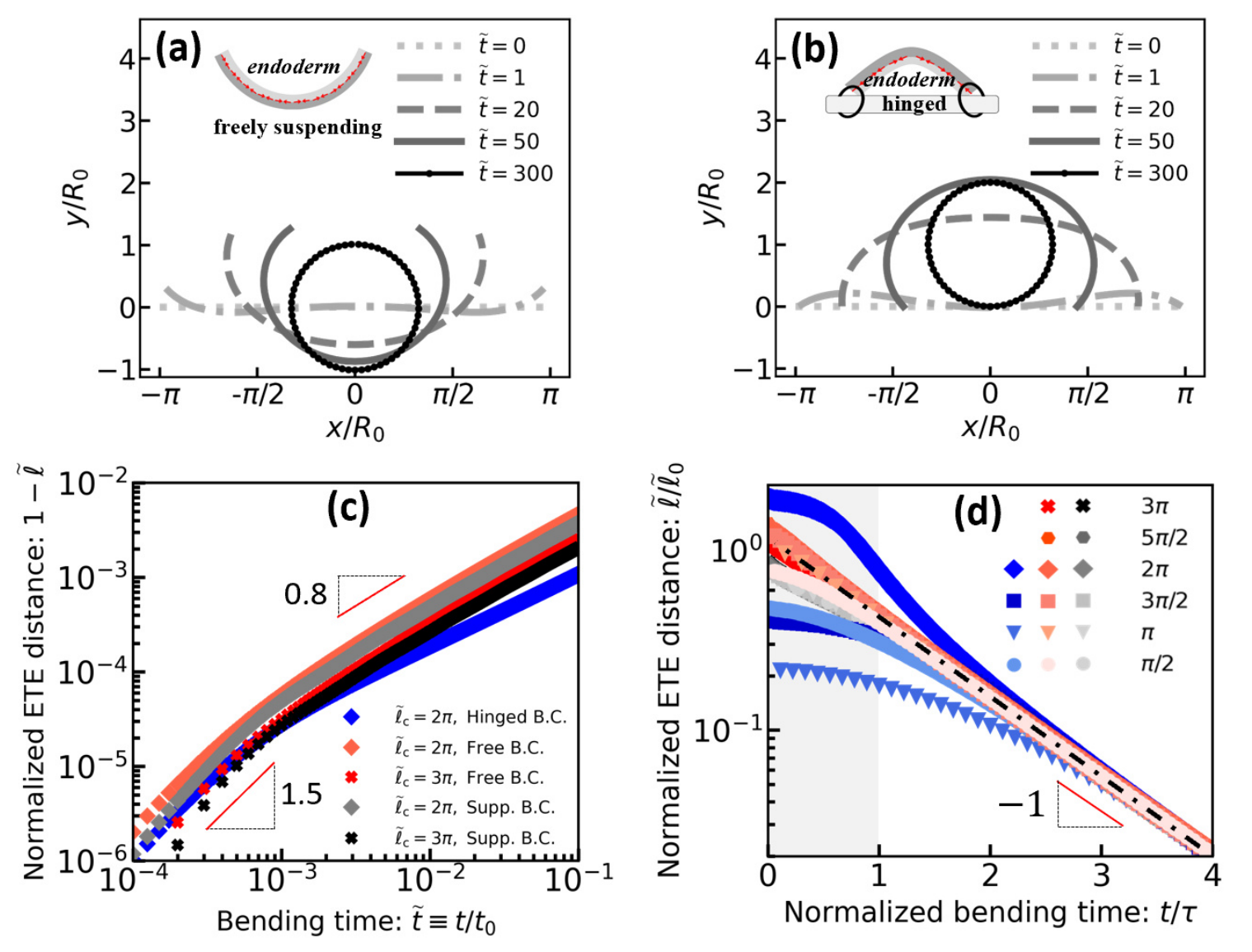}
  \caption {(a-b) Shape evolution of a rod-like Hydra fragment during its spontaneous bending in viscous fluids upon different boundary conditions: (a) free suspending rods (Free B.C., with $c_0>0$), and (b) rods (with $c_0<0$) on a solid surface with hinged boundary condition (Hinged B.C.). Here the contour lengths (with $\tilde{\ell}_{\rm c}\equiv \ell_{\rm c}c_0$) of the rods are both $\tilde{\ell}_{\rm c} = 2\pi$. (c-d) (color online) The temporal evolution of the normalized end-to-end (ETE) distance $\tilde{\ell}$ (defined in Eq.~(\ref{eq:BendDyn-results-regime1-ETE})) follows power-law relations in the initial bending stage (as shown in (c) by a log-log plot), and an exponential relation in the final bending stage close to the equilibrium bent state (as shown in (d) by a semi-log plot). Here different boundary conditions are marked by different colors (black for Supp. B.C., red for Free B.C., and blue for Hinged B.C.) and different (normalized) contour length are distinguished using different markers. $\tilde{\ell}_0$ is the interpolated (normalized) ETE distance for $t=0$ for rods upon different boundary conditions. The rod contour length $\tilde{\ell}_{\rm c}$ varies from $\pi/2$ to $3\pi$. 
  } \label{Fig:Schematic-BeadSpring3}
\end{figure}

Minimizing the Rayleighian
$\mathcal{R}=\dot{\cal F}_{\rm{b}}+\Phi$ with respect to $\dot{a}$ then gives
\begin{equation} \label{eq:minR3}
-\frac{1}{3} D_{\rm r}c_{0}^{2}+\xi_{\rm v} a \dot{a}\approx 0,
\end{equation}
from which we find
\begin{equation}\label{eq:at}
a \approx \left(\frac{2 D_{\rm r} c_{0}^{2}}{3\xi_{\rm v}}\right)^{1/2} t^{1/2},
\end{equation}
that is, the length of the bent part of the rod with non-zero curvature increase \emph{diffusively} with time scaling as $t^{1/2}$. Furthermore, the normalized ETE distance follows $1-\tilde{\ell} \sim a^3 \sim  t^{3/2}$. However, we would like to point out that the initial diffusive bending dynamics happens only in a very short period as shown in Fig.~3(a) in the main text. The bending dynamics slows down quickly when $t>10^{-3}t_0$ from $1-\tilde{\ell} \sim t^{3/2}$ to $1-\tilde{\ell} \sim  t^{0.9}$.

\subsection{Effects of boundary conditions and differential spontaneous curvature}\label{sec:BendDyn-BCs} 
\begin{figure}[htbp]
  \centering
  \includegraphics[width=0.8\columnwidth]{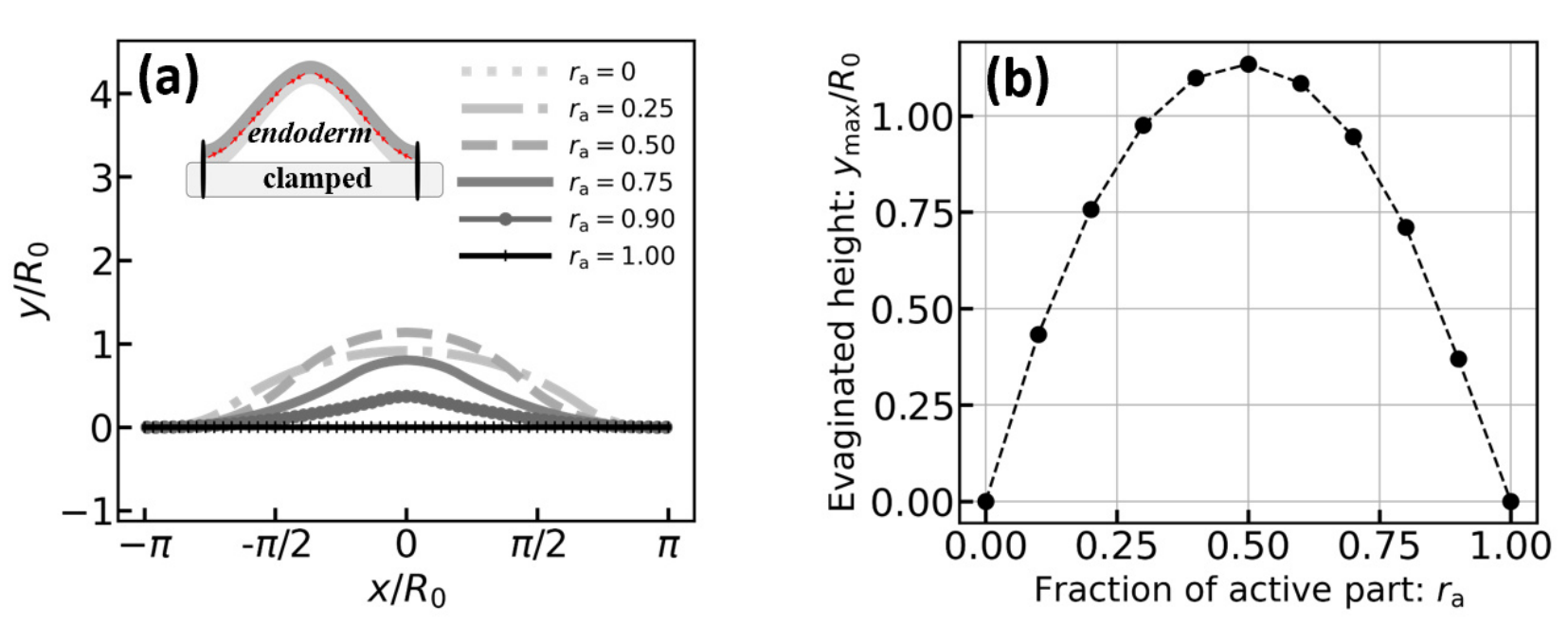}
  \caption {(a) Shape evolution of a Hydra rod-like fragment (of contour length $\ell_{\rm c} = 2\pi R_0$) during its spontaneous bending in viscous fluids on solid surface with clamped boundary condition. The rod has a differential $c_0$, in which the middle part of length $\ell_m$ (with $r_{\rm a}\equiv \ell_m/\ell_{\rm c}$) has a spontaneous curvature $-1/R_0$ while the remaining parts have $c_0=0$. In this case, the evaginated height (the maximum height, $y_{\rm max}$, of the equilibrium state) is plotted as a function of $r_{\rm a}$ in (b). 
  } \label{Fig:Schematic-BeadSpring4}
\end{figure}

For Hydra rods bending in viscous fluids, we use the bead-spring model mentioned above to explore the effects of different boundary conditions on 
the dynamics of their spontaneous bending. In this work, four different boundary conditions have been studied: (i) bending on non-sticky supporting solid surfaces (Supp. B.C., see Fig.~\ref{Fig:Schematic-BeadSpring2}(a)), (ii) bending of freely suspending rods (Free B.C., see Fig.~\ref{Fig:Schematic-BeadSpring3}(a)), (iii) bending on solid surfaces with hinged boundary condition (Hinged B.C., see Fig.~\ref{Fig:Schematic-BeadSpring3}(b)), and (iv) bending on solid surfaces with clamped boundary condition (Clamped B.C., see Fig.~\ref{Fig:Schematic-BeadSpring4}(a)).

For Hydra rods (with uniform spontaneous curvature $c_0$) bending upon the boundary conditions (i)-(iii), the temporal evolution of the end-to-end (ETE) distance (defined in Eq.~(\ref{eq:BendDyn-results-regime1-ETE})) is found to follow similar scaling relations. Initially, the bending of Hydra rods propagates \emph{diffusively} from the edges into the center, in which the bent length $a$ scales as $t^{0.5}$ and the normalized ETE distance $1-\tilde{\ell}\sim a^3$ scales as $t^{1.5}$ as shown in Fig.~\ref{Fig:Schematic-BeadSpring3}(c) and Fig.~3(a) in the main text. Whereas, when the bending is close to its final equilibrium (quasi-stable) shape, the end-to-end (ETE) distance $\tilde{\ell}$ decays exponentially with time towards its equilibrium value as shown in Fig. 3(b) of the main text and in Fig.~\ref{Fig:Schematic-BeadSpring3}(d). Moreover, the relaxational time shows the same scaling relation to the rod contour length as shown in Fig.~\ref{Fig:Schematic-BeadSpring2}(c). In addition, we have noticed that $1-\tilde{\ell}\sim t^{0.9}$ in the intermediate transition (slower sub-diffusive) regime for both free suspending rods and rods on supporting surfaces; however, the bending of rods upon hinged boundary conditions follows an even slower power-law as shown in Fig.~\ref{Fig:Schematic-BeadSpring3}(c).  

However, for Hydra rods with uniform $c_0$ upon clamped boundary condition (iv), the flat state is the energy-preferred equilibrium state as shown in Fig.~\ref{Fig:Schematic-BeadSpring4}(a) for the case of uniform spontaneous curvature with $r_{\rm a}=0$. In this case, we consider the case of non-uniform $c_0$, in  which the middle part of length $\ell_m$ (with $r_{\rm a}\equiv \ell_m/\ell_{\rm c}$) has a spontaneous curvature $c_0=-1/R_0$ while the remaining parts have $c_0=0$. In this case, the Hydra rod will buckle upward (see Fig.~\ref{Fig:Schematic-BeadSpring4}(a)) and the evaginated or invaginated height (the maximum height, $y_{\rm max}$, of the equilibrium state away from the solid surface) is plotted as a function of $r_{\rm a}$ in Fig.~\ref{Fig:Schematic-BeadSpring4}(b). An optimum fraction of the middle active part (\emph{i.e.}, the middle part with non-zero $c_0$) is found to be around $r_{\rm a}=0.5$. That is, for a given length of tissue fragments, the animal has to optimize the fraction of middle active part to achieve maximal depth of evagination or invagination.
 

\bibliography{HydraFolding}